\documentclass[a4paper,fleqn,usenatbib]{mnras}
\usepackage{multirow}
\usepackage{newtxtext,newtxmath}

\usepackage[T1]{fontenc}
\usepackage{ae,aecompl}
\usepackage{aas_macros}

\usepackage{graphicx,graphics}	
\usepackage{amsmath}	
\usepackage{amssymb}	
\usepackage{natbib}

\newcommand{\be}{\begin{equation}}
\newcommand{\ee}{\end{equation}} 
\newcommand{\bse}{\begin{subequations}}
\newcommand{\ese}{\end{subequations}} 
\newcommand{\bary}{\begin{eqnarray}}
\newcommand{\eary}{\end{eqnarray}}

\newcommand{\Msun}{\rm M_{\sun}}
\newcommand{\mseed}{m_{\rm seed}}

\newcommand{\K}{\rm K}

\newcommand{\Mpc}{\rm Mpc}

\newcommand{\lsim}{\mathrel{\hbox{\rlap{\lower.55ex\hbox{$\sim$}} \kern-.3em\raise.4ex\hbox{$<$}}}}
\newcommand{\gsim}{\mathrel{\hbox{\rlap{\lower.55ex\hbox{$\sim$}} \kern-.3em\raise.4ex\hbox{$>$}}}}

\newcommand{\REF}{Ref-L100N1504}

\newcommand{\RECAL}{Recal-L025N0752}

\newcommand{\soh}{$\nabla_{(\rm O/H)} $~}
\newcommand{\sohe}{$\nabla_{(\rm O/H)} $}

\defcitealias{schaye2015}{S15}
\defcitealias{crain2015}{C15}

\title[ Oxygen abundance gradients in the EAGLE discs]{The oxygen
  abundance gradients in the gas discs of galaxies in the EAGLE simulation}
\author[ Tissera et al.]{Patricia B. Tissera$^{1}$\thanks{E-mail:
patricia.tissera@unab.cl}, Yetli Rosas-Guevara$^{1,2}$,
Richard G. Bower$^{3}$, Robert A.  Crain$^{4}$,
\newauthor Claudia del P. Lagos$^{5,6}$, Matthieu Schaller$^{7}$,
  Joop Schaye$^{7}$, Tom Theuns$^{3}$
 .\\
$^{1}$Departamento de Ciencias Fisicas, Universidad Andres Bello,
700 Fernandez Concha, Las Condes, Santiago, Chile. \\
$^{2}$ Centro de Estudios de F\'isica  del Cosmo de Arag\'on, Plaza San
Juan 1, Planta 2, E-44001 Teruel, Spain.\\
$^{3}$Insitute of Computational Cosmology, Physics Department,
University of Durham , South Road, Durham DH1 3LE, UK.\\
$^{4}$Astrophysics Research Institute, Liverpool John Moores University, 146 Brownlow Hill, Liverpool L3 5RF, UK .\\
$^{5}$ International Centre for Radio Astronomy Research, University
of Western Australia, 35 Stirling Highway, Crawley, WA 6009,
Australia.\\
$^{6}$ ARC Centre of Excellence for All Sky Astrophysics in 3 Dimensions (ASTRO 3D).\\
$^{7}$Leiden Observatory, Leiden University, PO Box 9513, 2300 RA, Leiden, the Netherlands.}

\date{Accepted XXX. Received YYY; in original form ZZZ}

\pubyear{2015}

\begin{document}
\label{firstpage}
\pagerange{\pageref{firstpage}--\pageref{lastpage}}
\maketitle

\begin{abstract}
We use the EAGLE simulations to study  the oxygen abundance gradients  of gas discs in
 galaxies within the  stellar
mass range   $[10^{9.5}, 10^{10.8}]{\rm ~M_{\sun}}$ at
$z=0$. The estimated  median oxygen gradient
is $-0.011\pm 0.002$ dex~kpc$^{-1}$, which is shallower than
observed.
No clear trend between simulated disc oxygen gradient and galaxy stellar mass
is found when all galaxies are considered.
However, the oxygen gradient shows  
 a clear  correlation with  gas disc size so that shallower abundance slopes
 are found for increasing gas disc sizes.
Positive oxygen gradients are detected for  $\approx 40$  per cent of the
analysed gas discs, with a slight higher frequency  in low mass
galaxies. 
Galaxies that have quiet merger histories show a positive correlation
between oxygen gradient and  stellar mass, so that more massive galaxies tend to
have shallower metallicity gradients. At high stellar mass, there is a
larger fraction of 
rotational-dominated galaxies in  low  density regions. 
At low stellar mass,  non-merger galaxies show a large variety of
oxygen gradients and morphologies. The  normalization of  the disc oxygen gradients in non-merger
galaxies by the effective radius  removes the trend with stellar
mass. Conversely, galaxies that experienced mergers show a weak relation between oxygen
gradient and stellar mass. 
 Additionally,  the analysed EAGLE discs show  no clear dependence of
 the oxygen gradients on  local environment, in agreement with current observational findings.

\end{abstract}
\begin{keywords}galaxies: abundances, galaxies: evolution, cosmology: dark matter
\end{keywords}

\section{Introduction}
In the Local Universe, galaxy discs are known to have  oxygen abundance gradients in their interstellar medium (ISM)  that are (negative) steeper  with decreasing luminosity \citep[e.g.][]{lequeux1979,zaritsky1994} and  stellar mass \citep[e.g.][]{sanchez2013Califa, sanchez2014,ho2015}, i.e. the central abundances are larger than those in the outer regions.  However, recent results from the MaNGA survey show a weaker relation between metallicity gradients and stellar mass \citep{belfiore2017}, with a trend for smaller
and higher mass galaxies to have flatter  metallicity gradients than reported in  previous observations. This apparent tension between results obtained from different surveys and datasets
shows that we are still far from knowing the relation between metallicity gradients and stellar mass. With increasingly available data, it became clear 
that there is a large
variety  of metallicity gradients as a function stellar mass. 
This is also the case at high redshift where galaxies are reported to have  flat/positive \citep[e.g.][]{queyrel2012,troncoso2014} as well as  negative  \citep[e.g.][]{swinbank2012,jones2013,yuan2011} abundance
gradients \citep{carton2018}. An important issue to consider is that the determination of abundances is very sensitive to the adopted metallicity indicator and the spectral resolution \citep[e.g.][]{kewley2008,marino2013,mast2014}, so caution is always advised when comparing results from different surveys.

The formation of galaxies in the current cosmological paradigm is a complex and non-linear process  \citep{wr78}.
As gas cools and collapses within the  potential well of galaxies, stars form and evolve, injecting energy and/or chemical elements into the ISM at different stages 
of evolution. The new-born stars lock part of the enriched material into long-lived stars while the rest
is mixed  into the ISM. 
Within this context,  if galaxy formation proceeds in an inside-out fashion, negative metallicity gradients are naturally generated 
 \citep[e.g.][]{prantzos2000}. However, other mechanisms can alter or modify the metallicity gradients such as galactic fountains, gas inflows \citep[e.g.][]{amorin2012,molla2016}, galaxy-galaxy interactions 
\citep[e.g.][]{rupke2010,sillero2017}, ram-pressure, tidal stripping, amongst others. Together they  imprint chemo-dynamical patterns which can  be used to unveil the main processes responsible for their origin and evolution \citep[e.g.][]{freeman}.

Hydrodynamical simulations are a powerful tool to understand how metallicity gradients arise and evolve in disc galaxies, since the  chemical enrichment of
baryons can be tracked as galaxies are assembled in a cosmological context \citep[e.g.][]{mosconi2001,kg2003,koba2007,wiersma2009}.
The implementation of baryonic physics relies on subgrid modeling with  free parameters that need to be calibrated. Hence, the confrontation of
the simulated chemical abundances to observations not only provides a mean to understand their origin, but also a stringent route to set constraints on the models  \citep[e.g.][]{tissera2012,gibson2013,aumer2013}. 
Important results, which are complementary to those of numerical simulations, are provided  by analytical chemo-dynamical \citep[e.g.][]{matteucci1986,molla1997,chiappini2001}
and semi-analytical models \citep[e.g.][]{cora2006,delucia2014}. These approaches  yield negative metallicity gradients in discs if the gas cools and collapses  while conserving the angular momentum \citep{calura2012, molla2017}.

Previous studies using numerical simulations agree on the fact that an inside-out disc formation scenario leads  to negative metallicity gradients  \citep[e.g][]{calura2012,gibson2013,tissera2016}. 
 \citet{pilkington2012} find that the radial dependence of the star formation efficiency as a function of time sets the metallicity gradients for a given 
subgrid physics.
Enhanced SN feedback and high gas fraction have been shown to be related to the setting of positive metallicity slopes as gas can be ejected from the inner regions where the star formation activity tends
to be stronger \citep[e.g.][]{gibson2013,ma2017}.
\citet{tissera2016} report that their simulated gaseous discs show a correlation between metallicity gradients and stellar masses that agrees with the observational
results so that smaller galaxies tend to show a larger variety of metallicity slopes \citep{ho2015}.

Positive metallicity slopes (i.e. lower central abundances than in the outskirts)  in the ISM of discs are generally reported to be
asociated to galaxy-galaxy interactions in observations \citep[e.g.][]{molina2017} as well as in simulations \citep[e.g.][]{rupke2010,tissera2016,ma2017}.
Observations have shown that  the central metallicities of  galaxies in close interactions  are lower than those  in isolation, at least during certain stages of evolution \citep{kewley2006,ellison2008,dansac2008,dimatteo2009,kewley2010}.  Numerical simulations find that mergers and interactions are efficient mechanisms to trigger inward gas inflows \citep[e.g.][]{bh96,tissera2000,pedrosa2015,lagos2018}. This gas inflows can dilute the central gas metallicities if disc galaxies have initially negative metallicity gradients. In fact, hydrodynamical simulations  find a clear trend for galaxies in close pairs to exhibit lower central metallicity compared to galaxies in isolation during the first passage but then, the increase of
gas in the central regions triggers new starbursts which, in turn, enrich the ISM and regenerate negative profiles  \citep{perez2006, rupke2010, perez2011,sillero2017} .  

Until recently the simulated samples that have been used to study  metallicity gradients have been limited to those of galaxies selected from small cosmological volumes or zoom-in systems. The EAGLE project \citep[][]{schaye2015,crain2015} provides an opportunity to study a large simulated sample of metallicity gradients in a cosmological context.
 EAGLE galaxies have been shown to reproduce  a variety of observational relations at different scales  such as the colour-magnitude relations, the neutral gas content or the clustering properties \citep[e.g.][]{trayford2015,lagos2015,artale2017,crain2017}.
The mass-size relation is reproduced as part of the subgrid calibration of the EAGLE simulations at $z=0$ \citep{schaye2015}. These simulations are also able to reproduce the observed size evolution  \citep{furlong2015}.

Regarding disc formation, \citet{zavala2015} show that the stellar disc components in EAGLE preserve their specific angular momentum content as they formed
from gas that  conserved the specific angular momentum, in global agreement with the standard disc formation model \citep[][]{fall1980}. If bulges and discs are considered together then 
a net angular momentum loss is expected since stars in the spheroidal components formed from gas that lost  specific angular momentum \citep{obreja2013,pedrosa2015, stevens2017,lagos2018}.
 Due to the complex assembly history of
the structure in a hierarchical scenario, galaxies are subject to different processes that can redistribute the angular momentum content of  baryons such as
mergers, tidal stripping,  bar formation, outflows and  stellar migration. Therefore, it is the net balance between
losses and gains by the different galaxy components that finally matters  \citep[e.g.][]{sales2012,pedrosa2015,genel2018}. 

The relation between stellar mass, metallicity and star formation in EAGLE simulations has been analysed by \citet{derossi2017}. These authors find results in global agreement with observations, although a weaker redshift evolution of the mean abundances as a function of stellar mass is reported. The $\alpha$-abundances of massive  galaxies are analysed by \citet{segers2016}, showing that observed trends are globally reproduced.
\citet{mackereth2018} examine the elemental abundances of the disc stars of EAGLE's L$^\star$ galaxies, finding that abundance patterns similar to those revealed by the SDSS-III/APOGEE survey \citep[see e.g.][]{hayden2015} are uncommon. They conclude that the Milky Way's abundances and formation history are unlikely to be representative of the broader population of similarly massive, disc-dominated galaxies.

 In this paper,
we focus on the analysis of  the oxygen  abundance gradients  of the  star-forming  gas in the discs of central galaxies  identified in the 100 Mpc volume EAGLE simulation  (\REF) at $z=0$.
This simulation  includes the effects of  stellar and Active Galactic Nucleus (AGN) feedback.  Our EAGLE galaxy sample  comprises  $592$ galaxies with  discs that are  defined by at least 1000 baryonic particles. The analysed galaxies covered a wide range of  morphologies and are located  in different local  environments.

This paper is organised as follows. In Section 2 we describe the main characteristics of the simulations and the galaxy sample. In Section 3 we discuss the metallicity gradients and
SFR  of the  simulated disc galaxies and confront them with observations. In Section 4 we explore the physical mechanisms determining the metallicity gradients. 
Our main findings are summarised in Section 5. Appendix A.1. and  A.2 discuss the impact of numerical resolution and the impact of
varying the SN feedback.

\section{The EAGLE simulations}

We use cosmological simulations that are part of the  EAGLE project\footnote{ The reader is referred to http://eaglesim.org,
http://eagle.strw.leidenuniv.nl for a global description of the project, access to movies and images and to the database of galaxies are described in \citet{mcalpine2016}.}, which comprises a set of cosmological hydrodynamical simulations with variations in volume,  numerical resolution and galaxy formation subgrid models. Here, we only give a brief overview of the simulations and subgrid physics modelling. Extensive descriptions can found 
in \cite{schaye2015}  and \cite{crain2015}. 
The EAGLE simulations were carried out with a version of the parallel
hydrodynamic code  {\small GADGET-3} \citep{springel2005}. 
The modifications to the hydrodynamic solver  and time stepping are referred to  as {\small ANARCHY}  \citep{schaller2015}.

The initial conditions
 are consistent with  the Planck Cosmology parameters  \citep{planck2014}: $\Omega_\Lambda=0.693$, $\Omega_{\rm m}=0.307$, $\Omega_{\rm b}=0.04825$, $\sigma_8=0.8288$, $h=0.6777$, n$_{s}=0.9611$ and $Y=0.248$ where  $\Omega_\Lambda$, $\Omega_{\rm m}$ and  $\Omega_{\rm b}$ are the average densities of matter, dark energy and baryonic matter in units  of the critical density $z=0$,  $\sigma_8$ is the square root of the linear variance,  $h$ is the Hubble parameter ( $H_{o}\equiv h \,100 \rm km  s^{-1}$ ), $n_{s}$  is the scalar power-law index of the power spectrum of primordial  perturbations, and $Y$ is  the primordial mass fraction of  helium.
 The subgrid physics parameters was calibrated to reproduce the galaxy mass function at $z=0.1$ and the observed sizes of the galaxies today in the so-called reference model \citep{schaye2015}.  In addition,  other variations in the subgrid physics have been explored  as presented in  \cite{crain2015}. 

The largest EAGLE simulation (\REF) has a comoving cubic volume of  $100 ~\Mpc$  in linear extent.  The setup of the initial conditions provides a mass resolution of $9.7\times 10^6 \Msun$ for dark matter and  an initial mass of $1.81 \times 10^6 \Msun$ for baryonic particles. The gravitational calculations between particles are computed with  a Plummer equivalent softening length of 2.66  comoving kpc limited to a maximum physical size of 0.70 kpc. 
 In Table ~\ref{table:simulations}, we summarise  the main parameters of the simulated used in this work.

\subsection{Subgrid Physics} 

In this section, we summarise the main characteristics of the subgrid modeling. Full details and discussions are given by \cite{schaye2015}.
The radiative cooling and photoheating  implementations are described in \citet{wiersma2009a}. The radiative rates are calculated on an element-by-element basis for an ionised gas in equilibrium in the presence of an ionising UV/X-Ray background (model of \citealt{haardt2001}) and the Cosmic  Microwave  Background. Eleven elements (H, He, N, O, C, Ne, Mg, Si, S, Ca, and Fe) are tracked individually, with yields described by \citet{wiersma2009}. The radiative cooling and heating rates are computed with the software Cloudy \citep{ferland2013}. 
Collisional ionisation equilibrium  and non-ionising background are assumed prior to 
 reionization.

Star formation is implemented stochastically following the model of  \cite{schaye2008}, accounting for a metallicity-dependent star-formation density threshold,  $n^{*}_{\rm z}$ \citep{schaye2010}, of 
\begin{equation}
n^{*}_{\rm z} = 10^{-1} {\rm cm} ^{-3} \left( \frac{Z}{0.002}\right)^{-0.64},
\end{equation}
 to capture the  transition from a  warm,  atomic to a cold, molecular phase \citep{schaye2004}.  The star formation rate   
 reproduces the empirical Schmidt-Kennicutt law which is written in terms of a pressure law. A pressure floor is assumed as a function of density, $P \propto \rho^{\gamma\rm eff}$, for gas with
$\gamma_{\rm eff} = 4/3$. With this value of  $\gamma_{\rm eff}$,  the Jeans mass, and the ratio of the Jeans length
to the SPH kernel length are independent of density, preventing spurious
fragmentation due to a lack of resolution. Star-forming gas particles are stochastically selected  to become star particles which
represent a simple stellar population formed with  a \citet{chabrier2003}
Initial Mass Function. 

Stellar mass losses and metal enrichment due to stellar evolution are modeled as described in \citet{wiersma2009}.  The model tracks the nucleosynthesis and enrichment of the 11 elements mentioned above by three evolutionary channels:  (1) AGB stars, (2) supernova (SNe) type Ia and (3) massive stars and core-collapse SNe.  Stellar evolutionary tracks and
yields follow  \citet{portinari1998}, \citet{Marigo2001} and \citet{thie1993}. The abundance evolution is sensitive to the particular choice of yields.

The stellar feedback is treated stochastically, following  the thermal injection method described in \cite*{dallavecchia_schaye2012}.   The fraction of the available energy from
core-collapse SNe,  $f_{\rm th}$, injected into the ISM depends on the local metallicity and density as 
\begin{equation}
f_{\rm th} = f_{\rm th, \rm min} + \frac{f_{\rm th, \rm max} - f_{\rm th, \rm
min}}{1 + \left( \frac{Z}{0.1Z_\odot} \right)^{n_Z} \left( \frac{n_{\rm H, \rm
birth}}{n_{\rm H, 0}} \right)^{-n_{\rm n}}}, 
\label{eq:first}
\end{equation} 
where $Z$ is the metallicity, $n_{\rm H, \rm birth}$ is the density of the star particle 
inherited from the parent gas particle and $Z_\odot =0.0127$ is the solar metallicity. The parameters $n_{\rm H, 0}$  and $n_{\rm n}$ are calibrated to reproduce the galaxy mass function at $z=0.1$ and the sizes of galaxies today.  The amount of energy available is injected 30 Myr after the birth of the stellar population.  Neighbouring gas particles are  stochastically selected to be heated by  a fixed temperature difference of $\Delta T=10^{7.5}\K$.
  
Black holes  with  mass $\mseed = 1.48 \times 10^5 \Msun$ are placed in the centres of  haloes with mass greater than
$1.48 \times 10^{10} \Msun$,   using the BH seeding method  of \citet{springel2005}. Black holes can grow through merger events and gas accretion.  The accretion rates are computed as the modified Bondi-Hoyle accretion rate of \cite{rosas-guevara2015} and \cite{schaye2015}.  To modulate the Bondi- Hoyle accretion rate in high circulation flows, a viscosity parameter is introduced. The accretion rates are also limited by the Eddington rate.   Similar to the feedback from star formation, AGN feedback is treated stochastically and thermally. 
The AGN feedback parameters have been calibrated to reproduce the stellar mass galaxy function and the  scaling relation between stellar mass and the mass of the central black hole  for galaxies observed  in the Local Universe.

\begin{table*}
\caption{Main properties of the analysed EAGLE simulations and the values of the relevant subgrid parameters that differ between them. Columns show: (1) name, (2) box size,  (3) initial baryonic  particle number (the same number of dark matter particles was used), (4) initial baryonic and (5) dark matter particle mass, (6) comoving and (7) maximum proper gravitational softening, 8) $n_{\rm H,0}$  and (9) $n_n$ determine the characteristic density and the power law, respectively, of the density dependence of the available SN energy feedback from star formation (see Eq. \ref{eq:first}), (10) $f_{\rm th, \rm min}$ and (11) $f_{\rm th, \rm max}$ are, respectively, the asymptotic minimum and maximum  values of the fraction of energy feedback from star formation (see Eq. \ref{eq:first}). }
\begin{tabular}{|l|l|r|c|c|c|c|c|c|c|c|c|c|}
\hline
Name& L               &   $N$             &     $m_{\rm g}$         &  $m_{\rm DM}$ &  $\epsilon_{\rm com}$ & $\epsilon_{\rm prop}$   &   $n_{\rm H,0}$ & $n_n$ &  $f_{\rm th, \rm min}$ & $f_{\rm th, \rm max}$  \\
    & [$\rm cMpc$]    &                   & [$\rm M_\odot$]           & [$\rm M_\odot$] &     $[\rm ckpc]$ &          $[\rm pkpc]$ &      [cm$^{-3}$]        &    &  \\
\hline
Ref-L100N1504 & $100$  & $1504^3$  &  $1.81\times 10^6$& $9.70\times10^6$&  $2.66$     &  $0.70$     &  0.67        &  2/ln 10   &0.3& 3.0   \\
Ref-L025N0752    & $25$  & $752^3$  &  $2.26\times 10^5$& $1.21\times10^6$&  $1.33$     &  $0.35$  & 0.67         &  2/ln 10   &0.3& 3.0    \\
Recal-L025N0752 & $25$  & $752^3$  &  $2.26\times 10^5$& $1.21\times10^6$&  $1.33$     &  $0.35$ & 0.25        &  1/ln 10   & 0.3& 3.0    \\
WeakFB-L025N0376 & $25$  & $376^3$  &  $1.81\times 10^6$& $9.70\times10^6$&  $2.66$     &  $0.70$ & 0.67        &  2/ln 10   & 0.15& 1.5    \\
StrongFB-L025N0376 & $25$  & $376^3$  &  $1.81\times 10^6$& $9.70\times10^6$&  $2.66$     &  $0.70$ & 0.67        &  2/ln 10   & 0.6& 6.0     \\
\hline
\end{tabular}
\label{table:simulations}
\end{table*}

\subsection{The simulated galaxies}
In this work, we analyse the disc components of  central galaxies selected from the \REF ~simulation by using  the SUBFIND algorithm \citep{springel2005,dolag2009}.
The simulated galaxies are decomposed dynamically into a bulge and a disc component following the method and criteria described by \citet{tissera2012}. The stellar (gas) discs 
are defined by star (gas) particles with $\epsilon= J_{\rm z}/J_{\rm z,max}(E) > 0.5$,  where  $J_{\rm z,max}(E)$ is the maximum angular momentum along the
main axis of rotation, $J_{\rm z}$, over all particles at a given binding energy, $E$.  

Only those galaxies that have more than 1000 baryonic particles and  more than 100  star-forming gas particles in the disc components are  considered for the estimations of the metallicity profiles.
The final EAGLE  sample comprises 592 gas discs.

A set of properties of the selected galaxies are estimated such as the star formation rate (SFR) and specific star formation rate (sSFR), the disc stellar half-mass  radius (R$_{\rm eff}^{\rm stars}$),  the  half-mass radius of the gaseous discs (R$_{\rm eff}^{\rm gas}$) and the total half-stellar mass radius (R$_{\rm eff}^{\rm T}$). These parameters are estimated by considering particles within 1.5 $r_{\rm gal}$\footnote{The galaxy radius $r_{\rm gal}$ is defined as the radius that enclosed 83 per cent of the stellar mass of a system identified by the  SUBFIND algorithm \citep{springel2005}. This radius
mimics the isophotal 25 assuming a M/L$ = 1$ \citep[e.g.][]{tissera2010}.}. 
The stellar mass of the bulge and disc components identified by the dynamical decomposition described above are used to define the stellar disc-to-total mass ratio, D/T.  The EAGLE simulation \REF~yields galaxies with  large variety of galaxy morphologies. We analysed the metallicity gradients of the discs components regardless of galaxy morphology. When needed, we assume D/T~$=0.5$ to separate rotation-dominated from dispersion-dominated galaxies.

\begin{figure}
\resizebox{4.0cm}{!}{\includegraphics{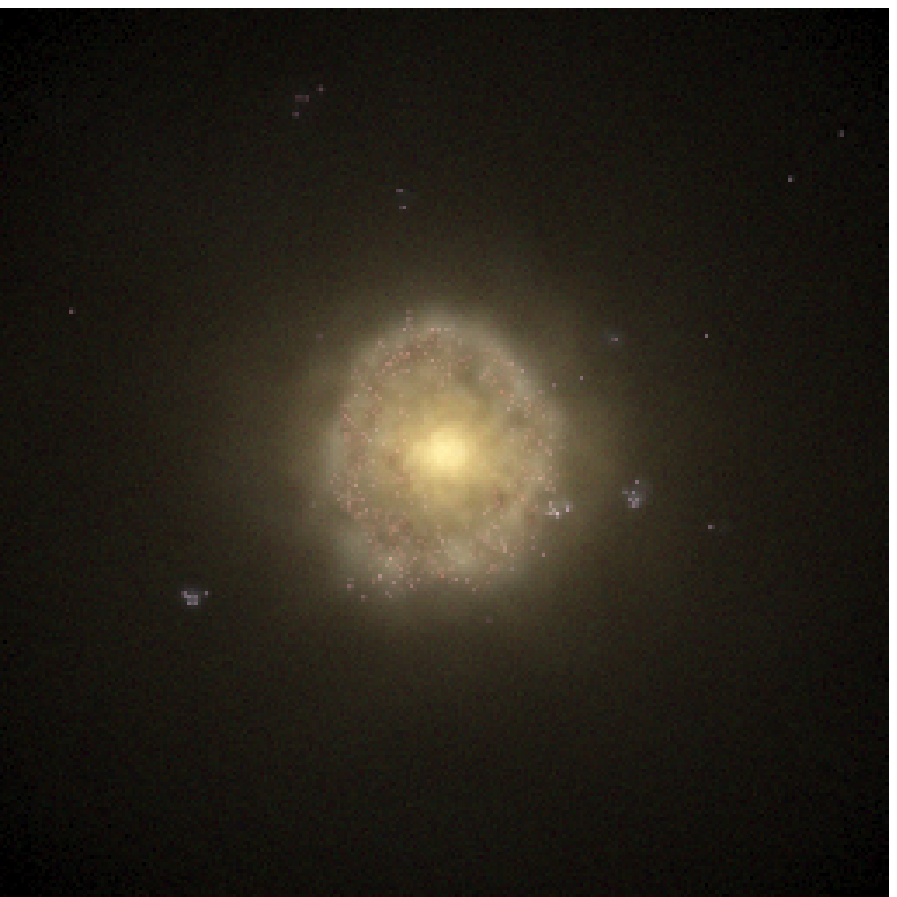}}
\resizebox{4.0cm}{!}{\includegraphics{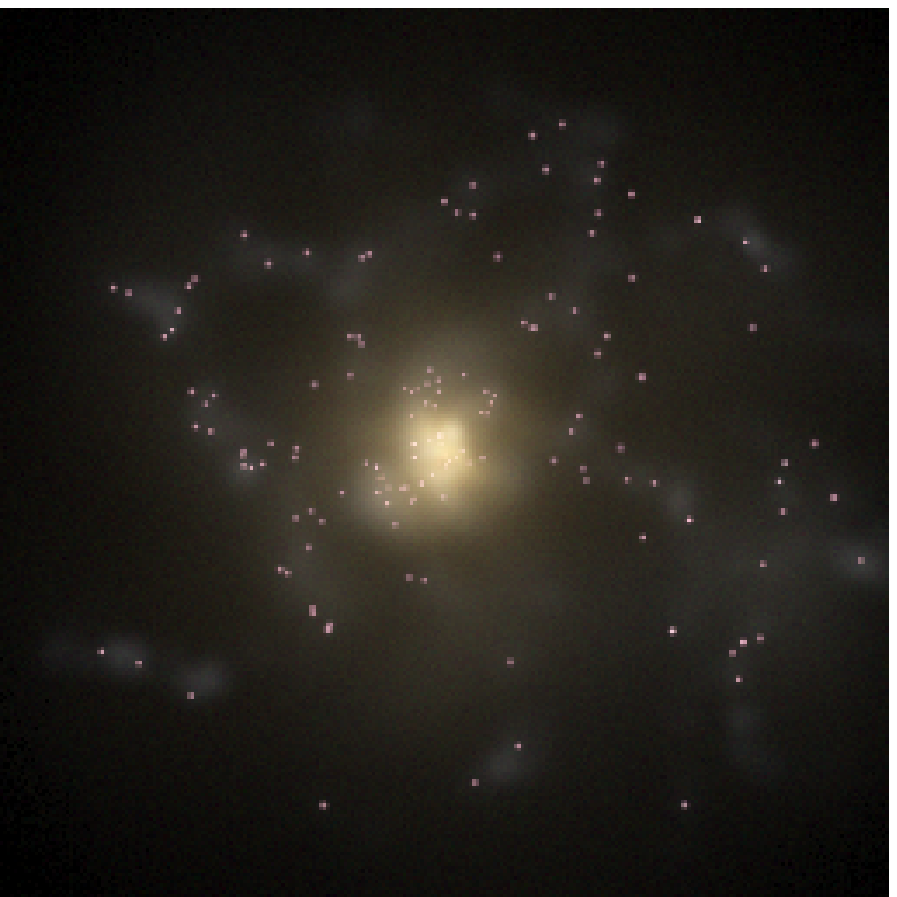}}\\
\resizebox{4.0cm}{!}{\includegraphics{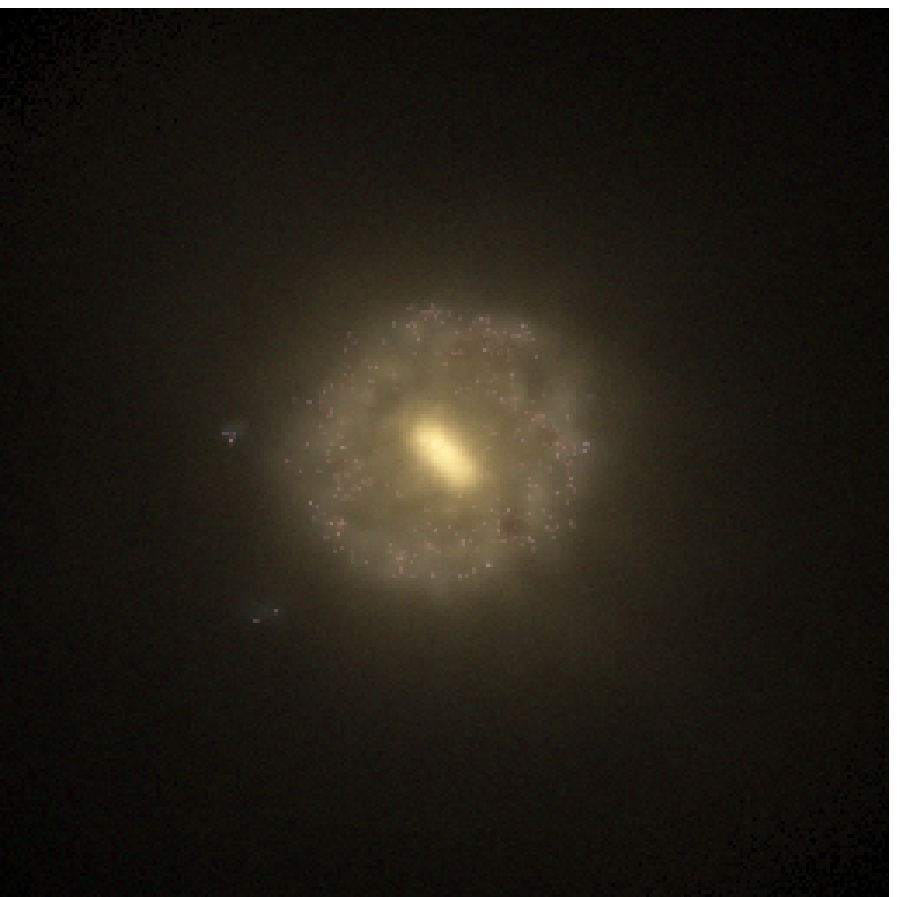}}
\resizebox{4.0cm}{!}{\includegraphics{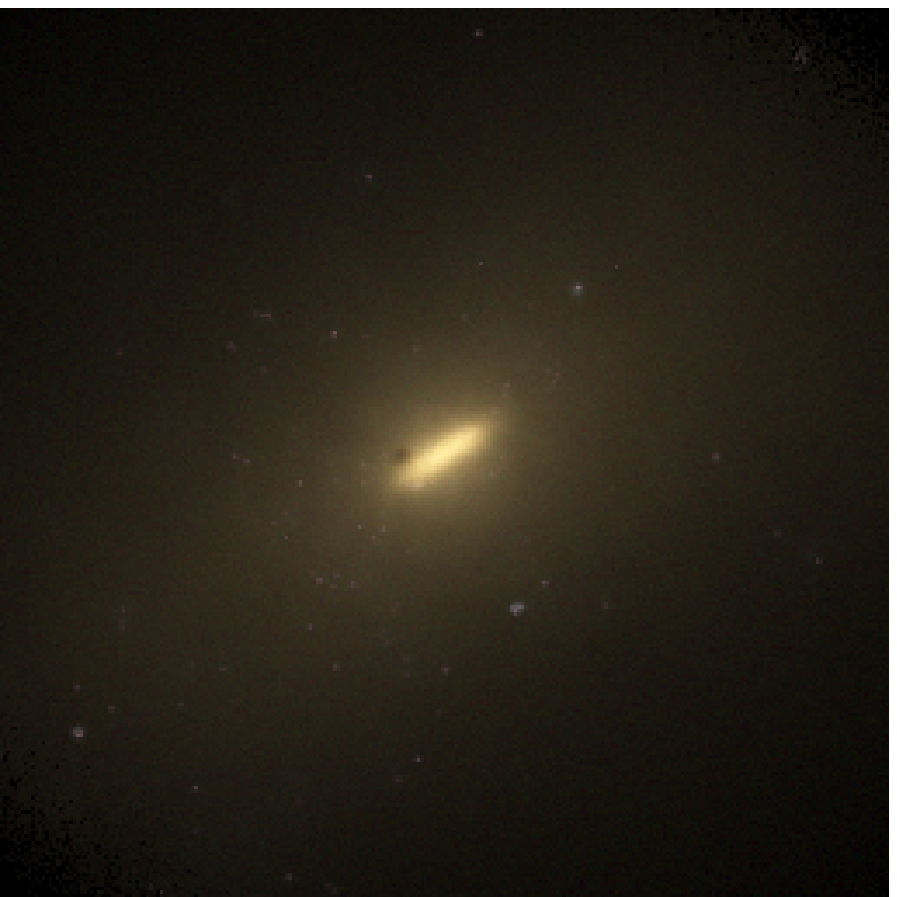}}\\
\resizebox{4.0cm}{!}{\includegraphics{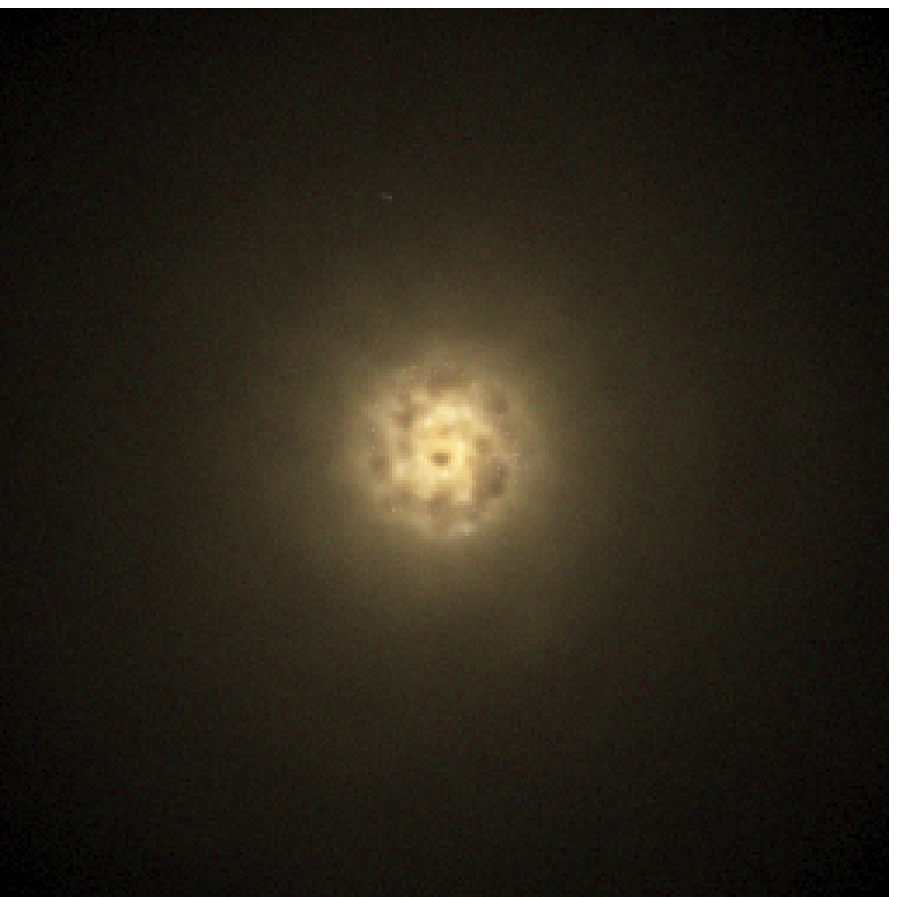}}
\resizebox{4.0cm}{!}{\includegraphics{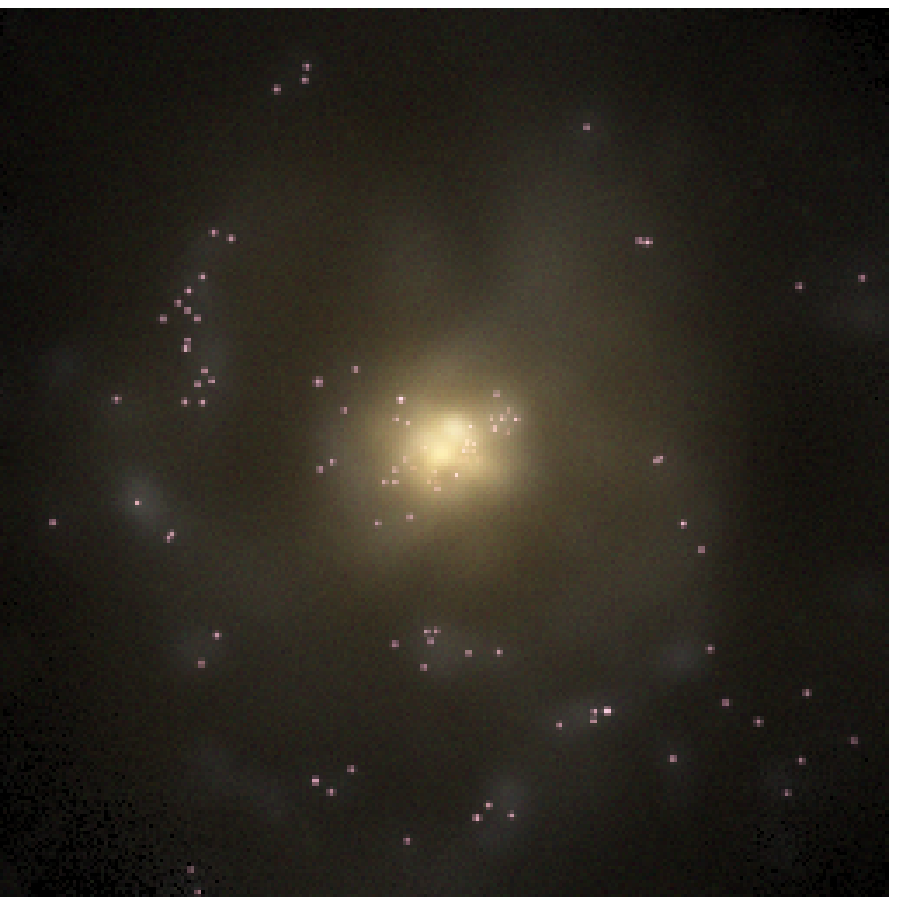}}
\caption{Face-on gri-band mock images of galaxies with negative (left panels) and positive (right panels)  gas-phase oxygen gradients of their disc components. Images are taken for the \REF~simulation from the EAGLE database \citet{mcalpine2016}.}
\label{images}
\end{figure}

To determine the metallicity profiles of the simulated gas  disc components, we use the oxygen abundances of the star-forming gas.  The  oxygen profiles are weighted by the star formation rate of the contributing regions 
to improve the realism of the comparison with observations.
A linear
regression is fitted within the radial range $[0.5, 2]$R$_{\rm eff}^{\rm stars}$ to estimate the slope (\soh)\footnote{Note that the metallicity dispersion at a given radius is quite large ($\sim $ dex), probably due
to the stochasticity related with the star formation and feedback modelling and inefficient metal-mixing within the simulated interstellar medium.}. 
For illustration purposes, we show  in Fig.~\ref{images}  examples of the gri-band mock images of  disc galaxies with positive (right panels) and negative (left panels)
metallicity gradients taken from the EAGLE database \citep{mcalpine2016} and as described in \citet{trayford2017}. 

A local galaxy density indicator is calculated  to analyse the metallicity gradients as a function of the environment using the count-in-cell (CC) method.  The simulated volume is divided into cubic volume cells of 0.5 Mpc on a side.  Then, the number of galaxies with stellar mass (M$_{\rm star}$) larger than $10^{9}$M$_{\sun}$ in each cell is counted, defining $N_{\rm CC}$. This method  provides similar results to the  fifth neighbour density estimators.  We note that similar trends are found when using the dark matter haloes instead.  
Low-density and high-density  environments are defined by adopting $N_{\rm CC} = 2$ as threshold. Most of
the analysed disc galaxies are located in low-density regions ($84$ per cent). The rest of these galaxies are distributed in groups with 3 to 23 galaxy members.
 
\begin{figure}
\resizebox{8.cm}{!}{\includegraphics{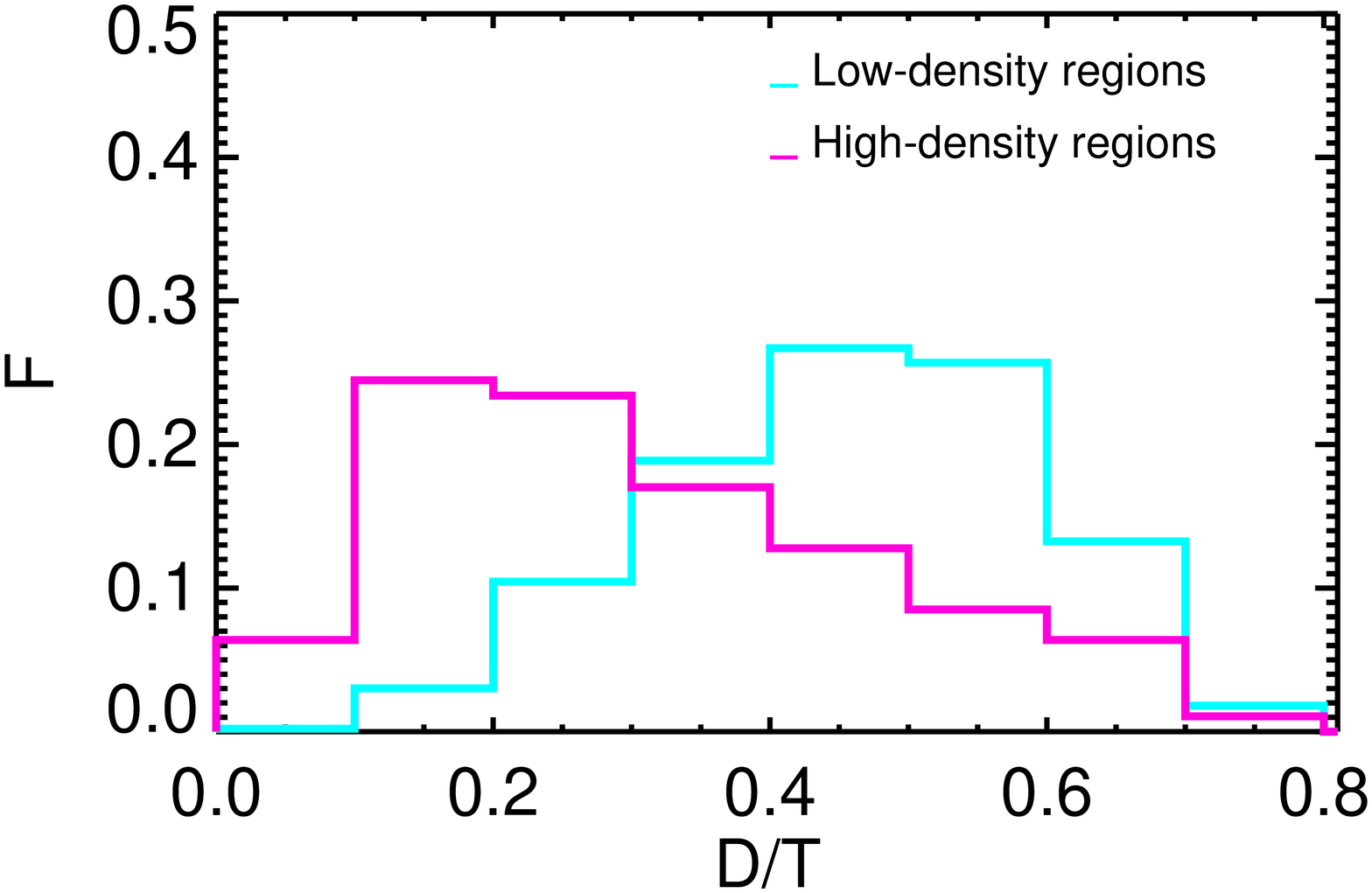}}
\resizebox{8.cm}{!}{\includegraphics{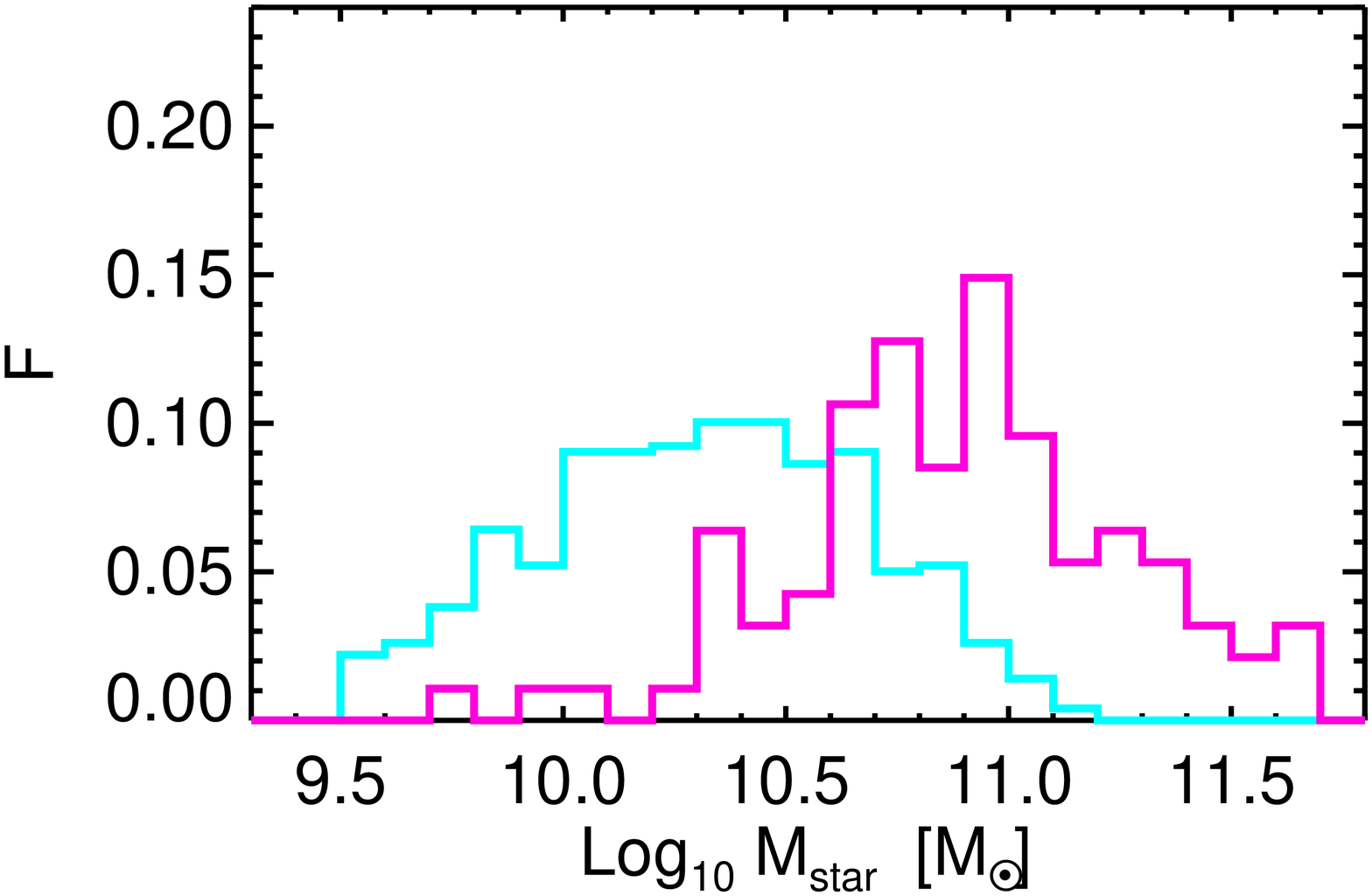}}
\resizebox{8.cm}{!}{\includegraphics{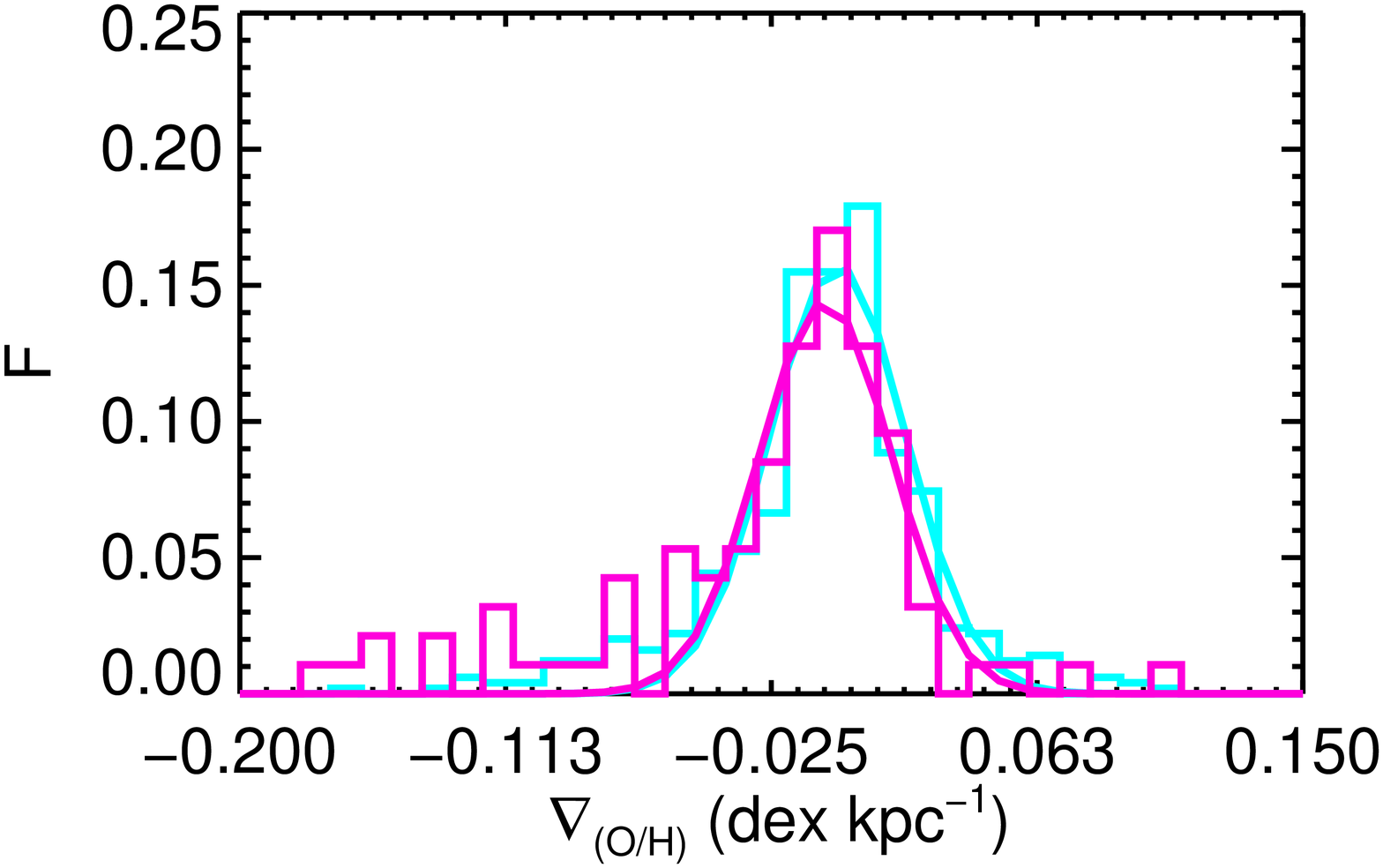}}
\caption{Distributions of D/T ratios (upper), stellar masses (middle)  and  metallicity slopes, \sohe, (lower) for the discs of the EAGLE galaxies in the two defined environments: low-density 
(N$_{\rm CC} \leq 2$; cyan line),
  and high-density (N$_{\rm CC} > 2$; magenta lines) regions. In the lower panel, the best Gaussian fits to the both simulated distributions are also shown.}
\label{dt_environ}
\end{figure}

As can be seen in the upper panel of Fig. ~\ref{dt_environ}, the EAGLE galaxies are able to
reproduce  the observed morphology-density dependence with most of the dispersion-dominated galaxies located in groups \citep{dressler1980}. 
Recall that we analyse the \soh of gas discs in central galaxies of different morphological types (i.e. different D/T ratios).

The middle panel of Fig.~\ref{dt_environ} displays the distribution of the galaxy stellar masses in the two defined local environments. As expected, there is a trend for  higher stellar mass galaxies to be located
in the high-density environments while  less masive ones tend to be in  low-density regions. Unfortunately, there is only a narrow stellar mass range that is sampled simultaneously in both local  environments. Hence, our estimations should be taken only as indicative.

In the bottom panel of Fig.~\ref{dt_environ}, the \soh distributions are shown as a function of local
environment. There is no significant statistical difference between them, although a small displacement of the \soh to 
more positive values can be seen in
low density regions. Gaussian fits to the \soh distributions yield mean values of $-0.007$  dex~kpc$^{-1}$  and $-0.003$ dex~kpc$^{-1}$  (widths of 0.022  dex~kpc$^{-1}$) for the high-and-low-density subsamples, respectively.
The bootstrap medians are $-0.009\pm 0.002$ dex~kpc$^{-1}$ and $-0.026 \pm 0.005$ dex kpc$^{-1}$ for  galaxies in  low-and-high-density regions, respectively. The median \soh for the whole disc samples is  $-0.011\pm 0.002$  dex~kpc$^{-1}$.  The differences between the medians and the Gaussian fit estimations are due to the existence of a tail towards more negative \soh in  high-density regions, which are not well represented by a Gaussian fit.

\section{Disc metallicity gradients}

\begin{figure}
\resizebox{9cm}{!}{\includegraphics{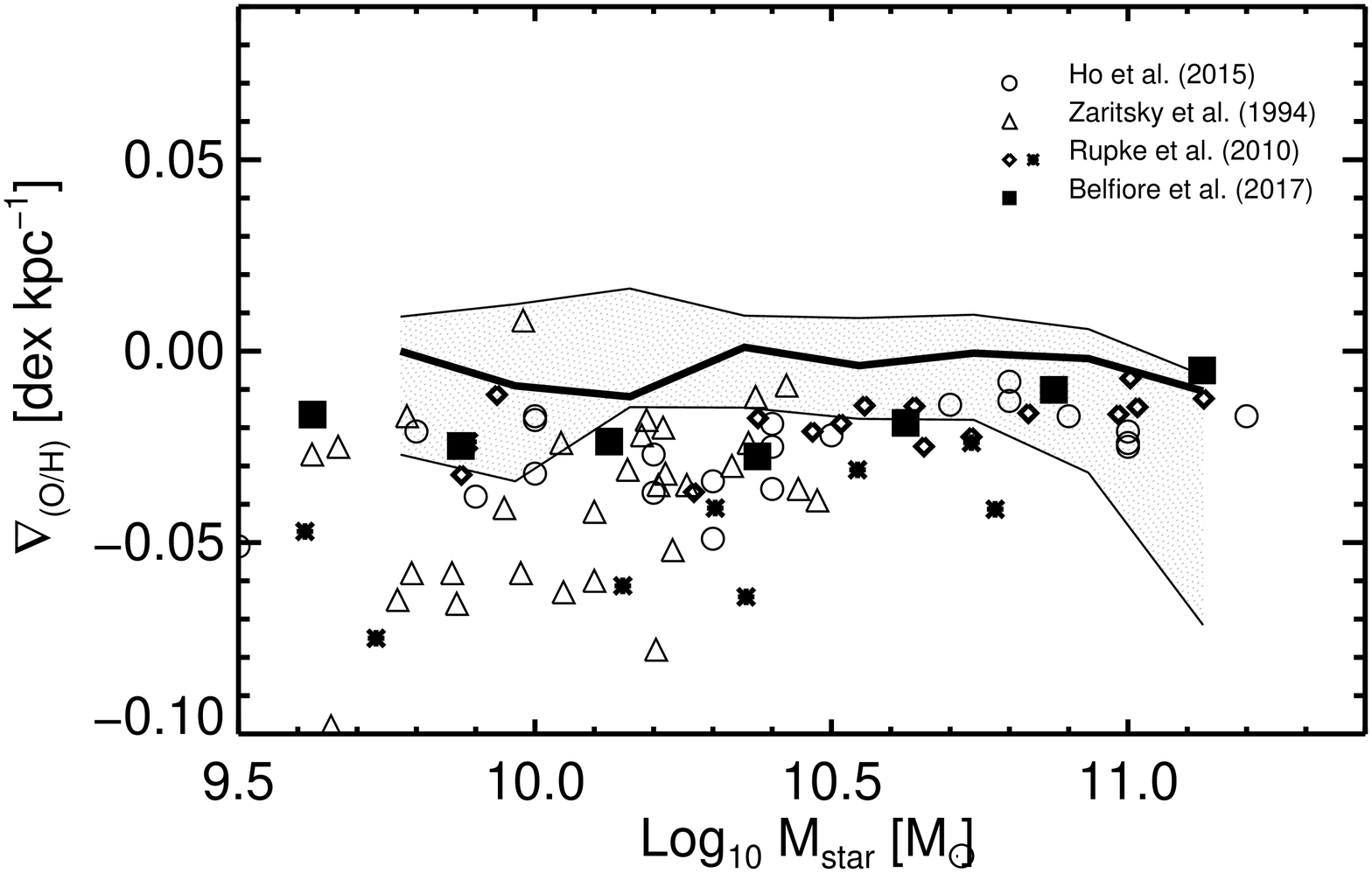}}
\caption{Median oxygen abundance slopes for the star-forming gaseous disc components as a function of stellar mass of the selected EAGLE galaxies (black  line).
The  shaded areas  are defined by the first and third quartiles. For comparison, observational results from  \citet[][open triangles]{zaritsky1994}, \citet[][open diamonds]{rupke2010}, \citet[][open circles]{ho2015} and \citet[][filled squares]{belfiore2017} are included.}
\label{gradients}
\end{figure}

In Fig.~\ref{gradients}, we show \soh of the gas discs as a function of the stellar mass of  the selected EAGLE galaxies.
 As can be seen, there is no clear trend between the \soh and M$_{\rm star}$. The simulated median \soh are  slightly above the observational values reported for
galaxies in the Local Universe (black symbols). In particular, we estimate an offset of $\sim -0.02 $ dex~kpc$^{-1}$  at $\sim 10^{10.5}{\rm M_{\sun}}$ with respect to the MaNGA results reported by \citet[][see also Appendix A.1 for a discussion on the robustness of this result against numerical resolution]{belfiore2017}.
The simulated relation shows  more negative 
metallicity gradients  at the low and the higher stellar-mass ends where the scatter increases. 
From Fig.~\ref{gradients}, we can see that there is a significant fraction of oxygen profiles with positive \soh in the EAGLE discs (i.e. lower central abundances). We find  42 per cent and 30 per cent of
positive \soh in galaxy discs located in low-and-high density regions, respectively.
 Discs with positive oxygen slopes also contribute to establish  a weak mean trend with stellar mass by increasing the scatter of the \soh at a given stellar mass. This trend seems  at odds with some observations \citep[e.g.][]{zaritsky1994,rupke2010,ho2015}. However,  recent results by \citet{belfiore2017}  from MaNGA survey show a significant contribution of positive abundance slopes. We note that these authors also find an increase of the median \soh for low and high stellar-mass galaxies that is not detected in the EAGLE discs.

\begin{figure*}
\resizebox{8.5cm}{!}{\includegraphics{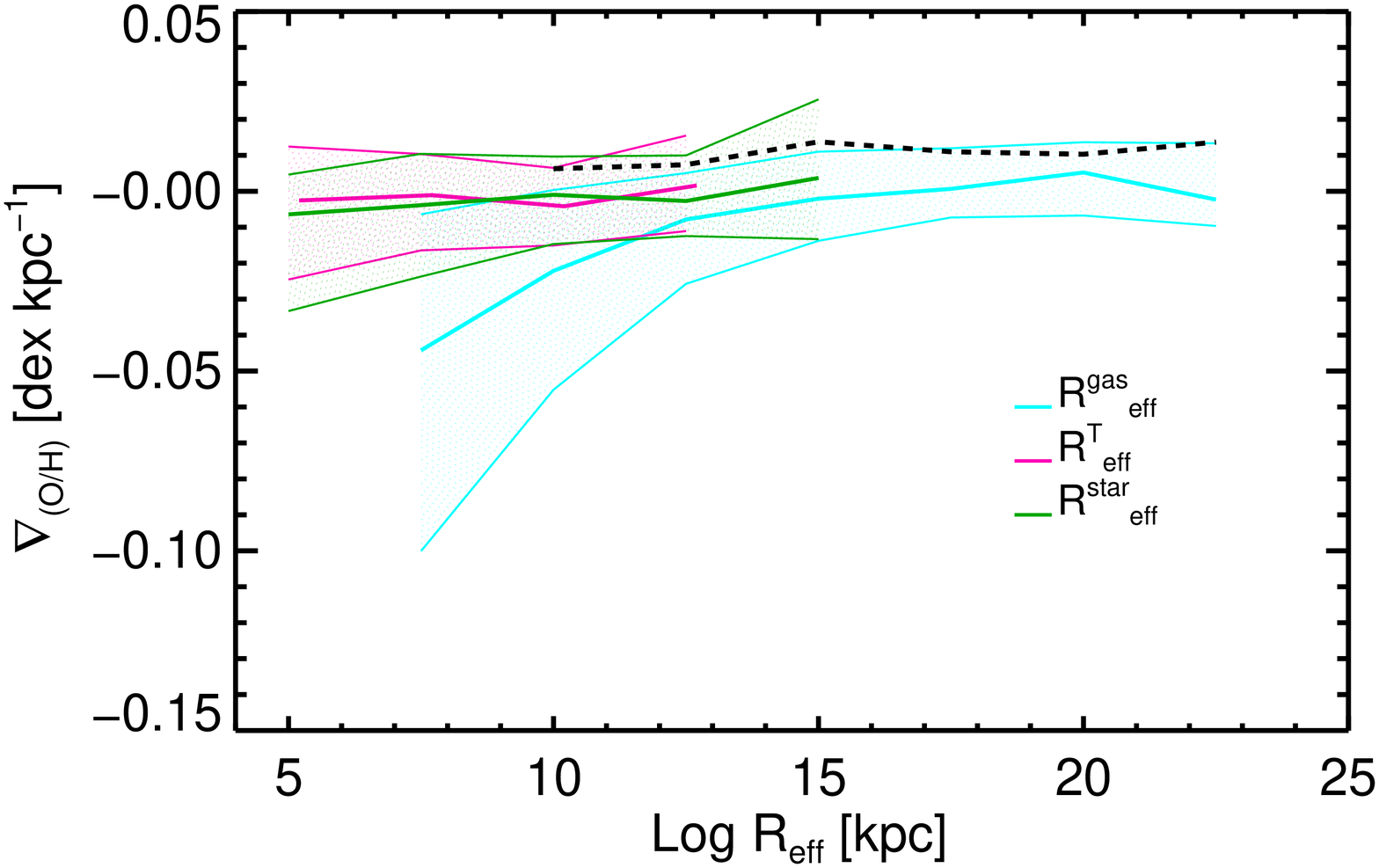}}
\resizebox{8.5cm}{!}{\includegraphics{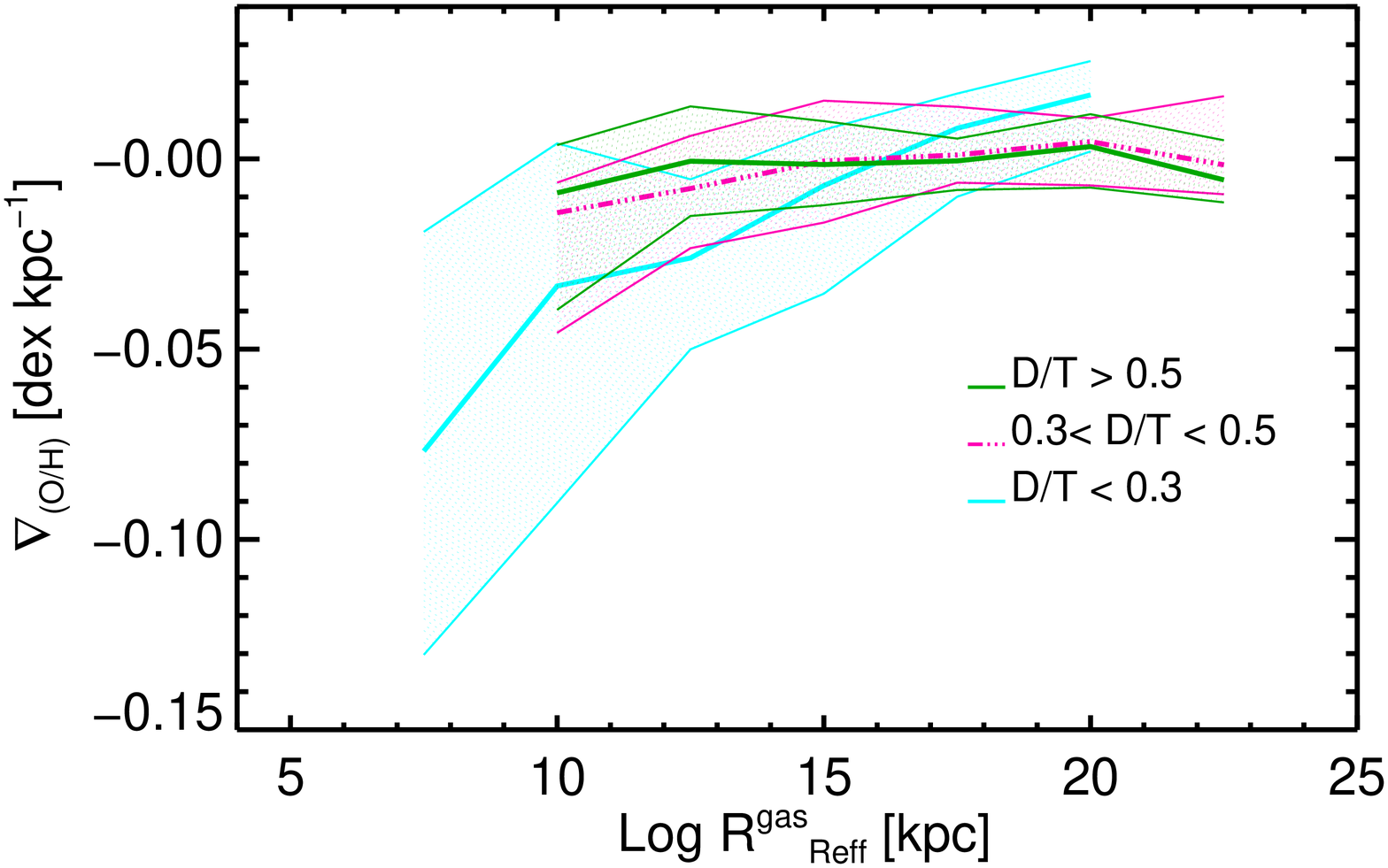}}
\caption{Median oxygen abundance slopes for the star-forming gaseous disc components as a function of  disc size of the selected EAGLE  galaxies. Left panel: Median values are shown as a function of  stellar (R$_{\rm eff}^{\rm star}$) and gas  (R$_{\rm eff}^{\rm gas}$) half-mass radius (green line and  cyan lines, respectively) and the galaxy stellar half-mass radius (R$_{\rm eff}^{\rm T}$; magenta line). The  shaded areas are defined by the first and third quartiles. Additionally, the relation for galaxies with positive metallicity gradients in the disc components is shown (black dashed line).
Right panel: Median oxygen abundance slopes as a function of R$_{\rm eff}^{\rm gas}$ for galaxies grouped according to D/T ratios. }
\label{gradients_reff}
\end{figure*}

In the left panel of Fig.~\ref{gradients_reff}, we show \soh as a function of galaxy size, considering  R$_{\rm eff}^{\rm T}$ (magenta line), R$_{\rm eff}^{\rm star} $ (green line) and R$_{\rm eff}^{\rm gas}$ (cyan line).  As can be appreciated from the figure,
there is a clear correlation between   \soh  and  R$_{\rm eff}^{\rm gas}$ so that  more negative \soh are found in galaxies with more compact gas discs.
The scatter of the relation increases for decreasing disc size.
 Conversely, discs with positive \soh show no trend with R$_{\rm eff}^{\rm gas}$ (black dashed line). Hence, the correlation is set only by discs with negative \sohe.
This trend can be interpreted as the results of  an inside-out disc formation  where the systems grow outwards while the star formation  moves progressively to the disc outskirts, enriching the ISM and flattening the abundance profiles. The larger the gas discs, the flatter the metallicity gradients \citep{prantzos2000}, if the discs are not disturbed (by a bar or mergers, for example).
Conversely, there is no clear correlation between \soh and  R$_{\rm eff}^{\rm T}$ or R$_{\rm eff}^{\rm star} $.

By using the D/T ratio as an indicative of galaxy morphology, we find a trend for discs in dispersion-dominated galaxies (D/T$<0.3$) to show a 
clear trend  between  \soh and   R$_{\rm eff}^{\rm gas}$. As can be seen from the right panel of Fig.~\ref{gradients_reff}, the smaller the galaxy, the steeper
negative \soh it has.
 Disky galaxies  (D/T$>0.3$) show a shallower relation. This suggests that the latter have been able to enrich more efficiently the ISM  while the formers have less
evolved disc components.

\section{What physics determines metallicity gradients?}

To explore how the \soh are determined in the EAGLE discs, we analyse possible dependences of \soh on other galaxy properties
as a function of stellar mass.  For this purpose, the original sample is divided into a high and a low stellar-mass subsamples, adopting  M$_{\rm star}=10^{10.5} {\rm M_{\odot}}$ as a threshold. Within each mass subsample, discs are grouped according to the disc \soh (i.e. negative or positive values). We find a  slightly larger fraction of positive \soh in the low-mass subsample, $0.43 \pm 0.06$, than in the high-mass one, $ 0.33 \pm 0.06$ (bootstrap errors are given).

\begin{figure}
\resizebox{6.5cm}{!}{\includegraphics{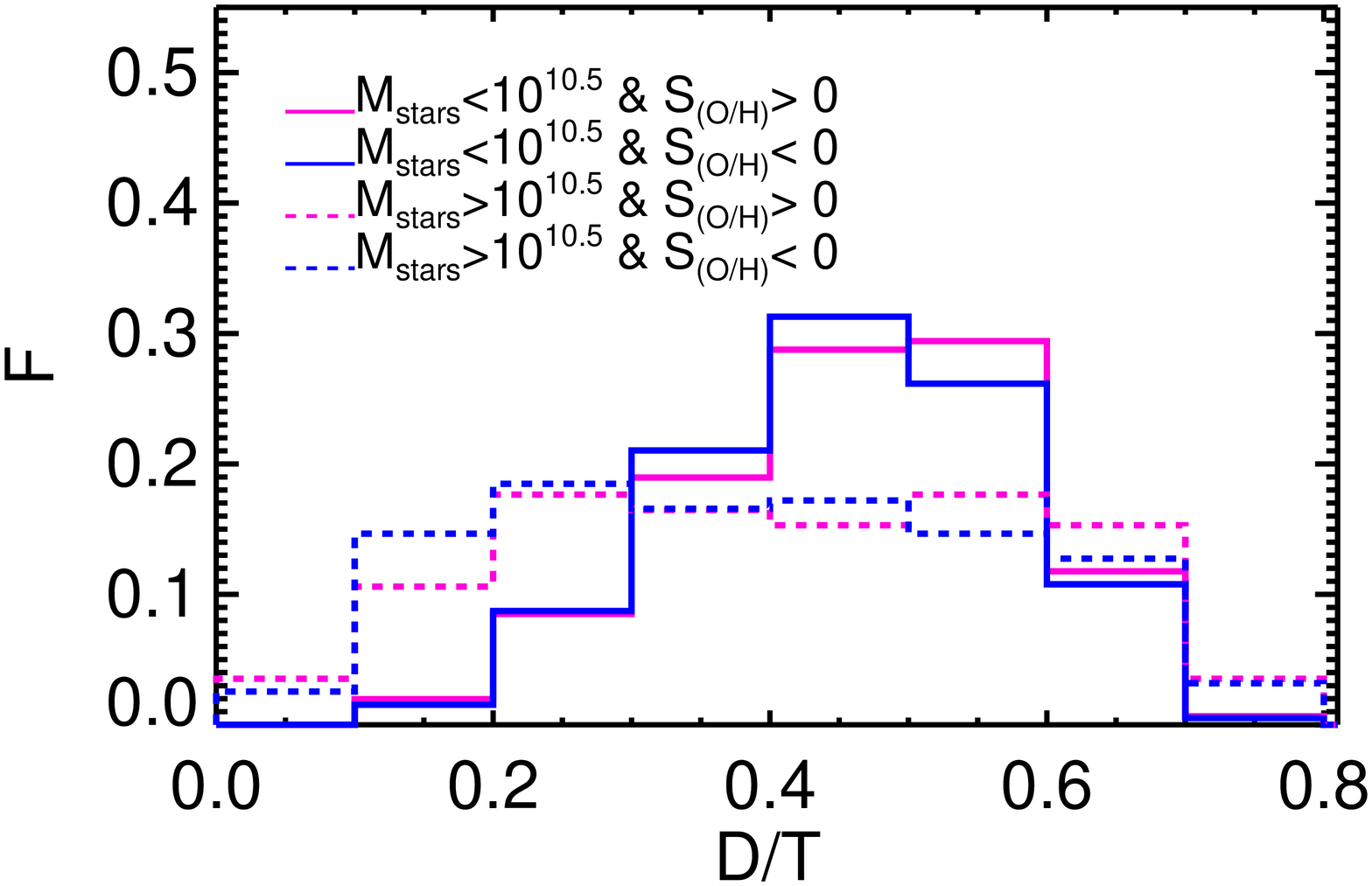}}
\resizebox{6.5cm}{!}{\includegraphics{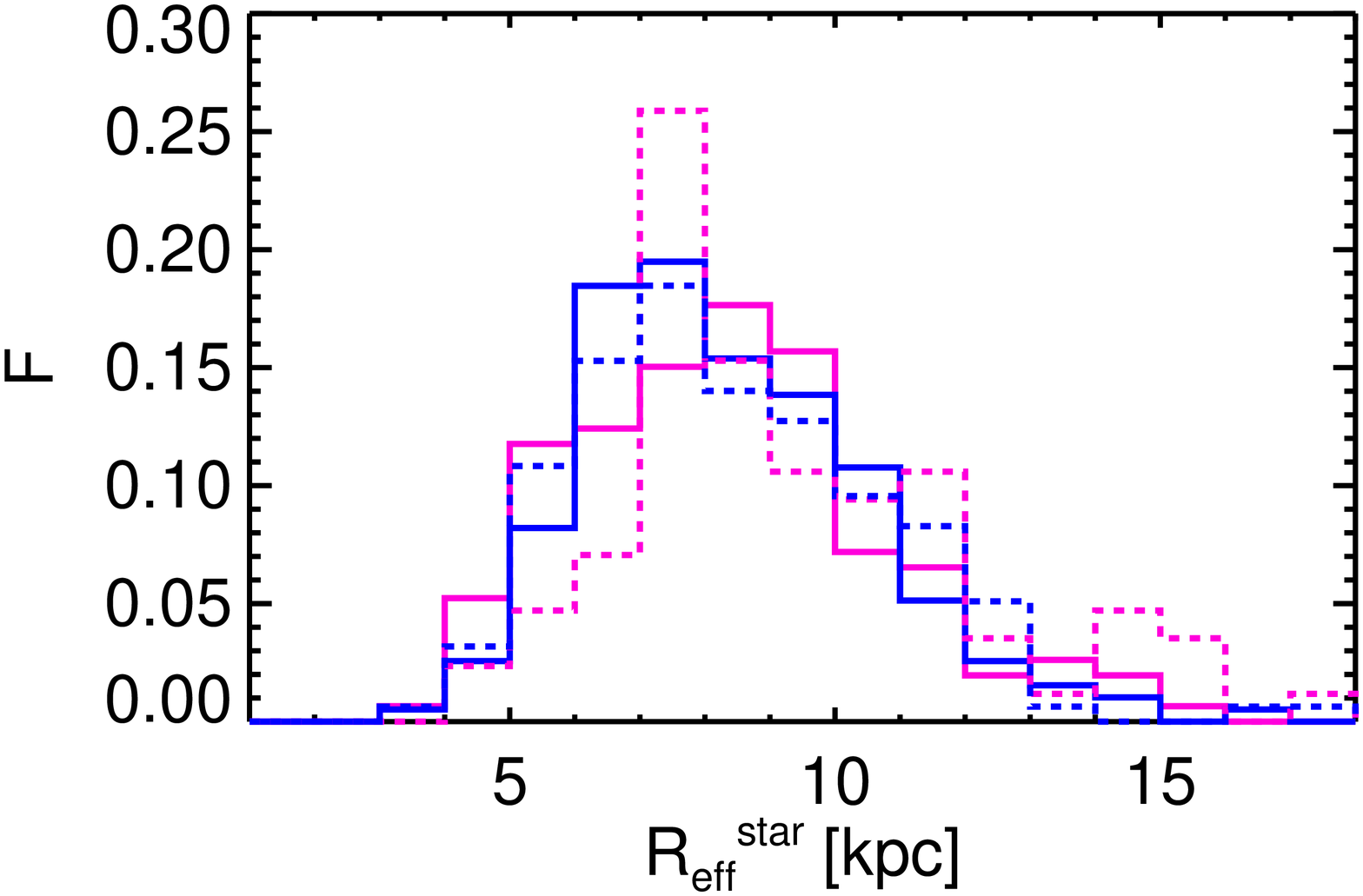}}
\resizebox{6.5cm}{!}{\includegraphics{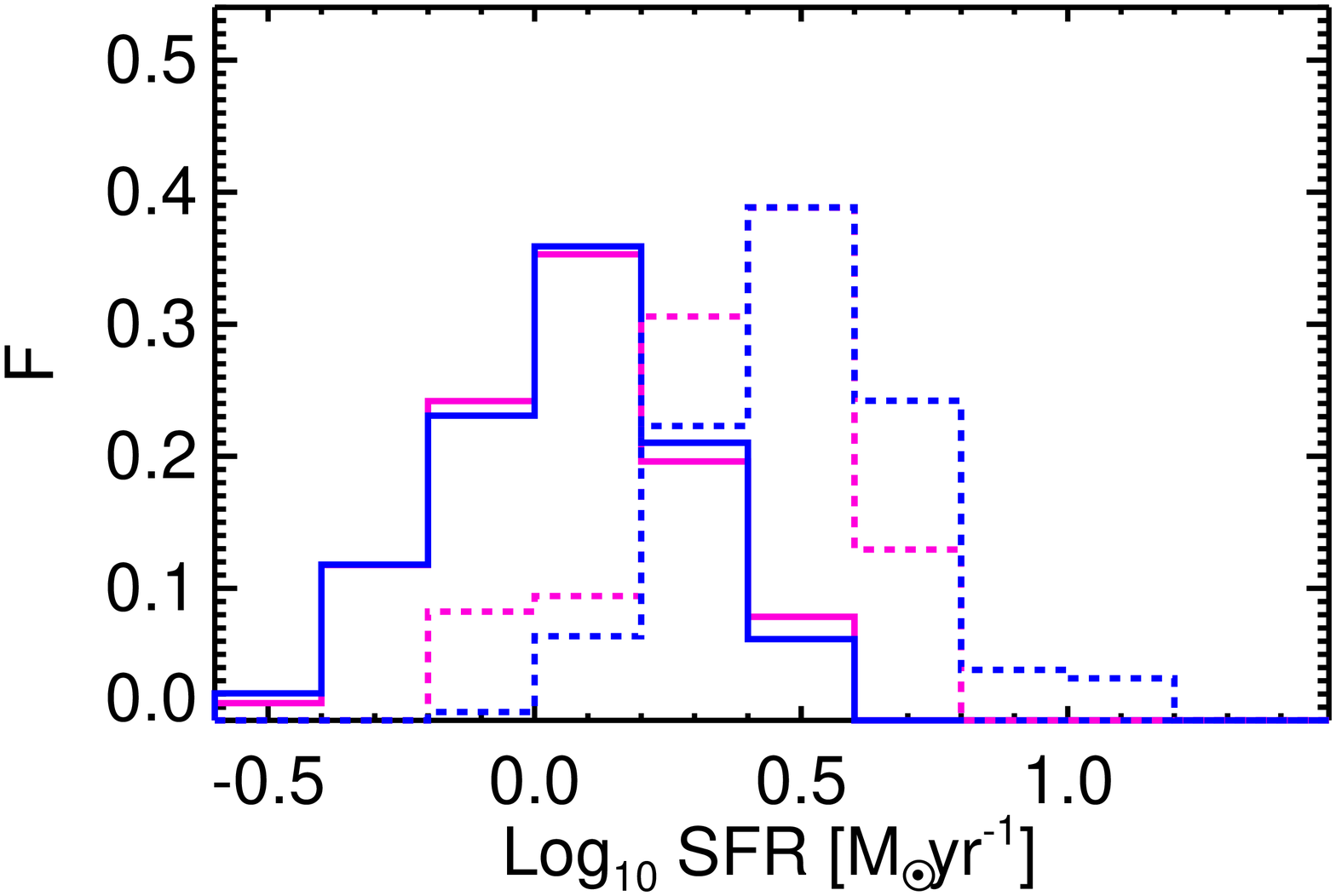}}
\resizebox{6.5cm}{!}{\includegraphics{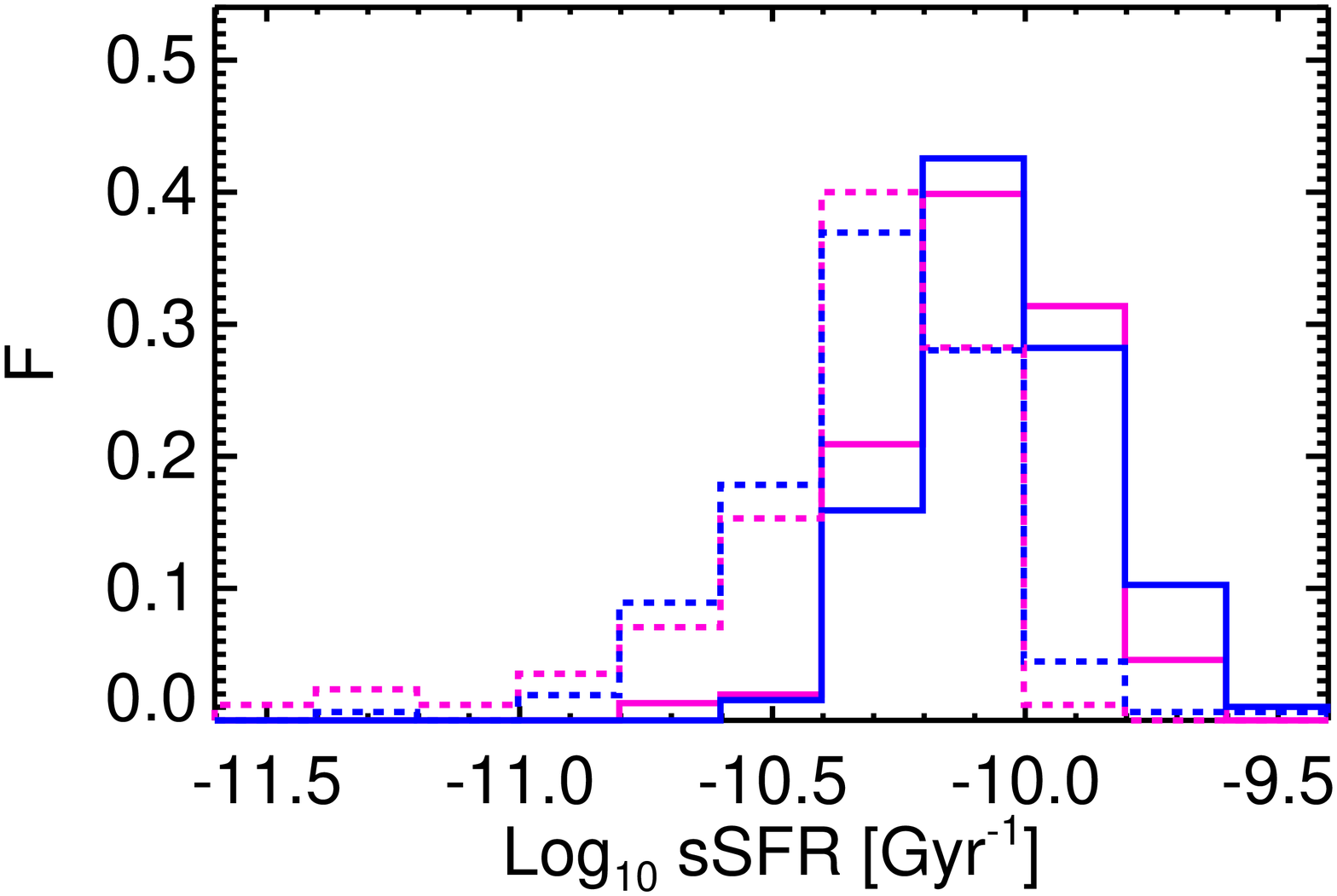}}
\caption{Stellar mass-weighted distributions of D/T ratios (first panel), $R_{\rm eff}^{\rm stars}$ (second panel),   SFR  (third panel) and sSFR (fourth panel) for $M_{\rm star} < 10^{10.5} {\rm M_{\odot}}$ (solid lines) and $M_{\rm star} > 10^{10.5} {\rm M_{\odot}}$  (dotted lines) galaxies with positive (magenta lines) and negative (blue lines) oxygen abundance gradients. }
\label{histoprop}
\end{figure}

In Fig. ~\ref{histoprop}, the distributions of D/T ratios,  R$_{\rm eff}^{\rm star}$, SFRs and sSFRs  for discs  with positive (magenta lines) and
negative (blue lines) \soh in  the low (solid lines) and high (dashed lines)  mass subsamples  are shown.
As can be seen from the first panel, there is a larger fraction of high mass galaxies with D/T~$ <0.3$ while low stellar-mass galaxies
 show a clear peak at D/T~$\sim 0.5$. The D/T distribution of massive galaxies is more uniform.

The second panel of  Fig. ~\ref{histoprop}  shows the  distributions of   $R_{\rm eff}^{\rm stars}$ which are consistent with 
more massive galaxies having slightly larger discs $R_{\rm eff}^{\rm star}$  \citep{furlong2015}. However, no clear differences in the size distributions are found for galaxies with positive or negative \soh  in a  given stellar-mass interval (according to a KS test).
The  third panel of   Fig. ~\ref{histoprop}  shows the  expected trend for massive galaxies  to have higher SFRs. The distributions of massive galaxies with positive \soh 
are slightly displaced towards lower SFR compare to those with negative \sohe. Smaller galaxies with positive and negative \soh  show no differences in the sSFR
 distributions as can be seen in the fourth panel.

The KS tests performed over these distributions provide no statistical differences, with only a weak signal for massive galaxies with positive \soh to be less active.
Recall that, as shown in Fig.\ref{gradients_reff}, there is a strong dependence of \soh on  $R_{\rm eff}^{\rm gas}$ 

\subsection{Disc assembly and merger history}

In order to get further insight in the origin of the characteristics of the abundance profiles, in this section,  we explore a set of properties  related to the history of disc assembly  the EAGLE discs.  

First, we calculate the cumulative fraction of the stellar mass formed as a function of stellar age for  discs in each of the subsamples used in  the previous section.  
We define $T^{30}$ as the lookback time (Gyr) when the last 30 per cent of the stellar populations was formed in each disc. This parameter aims at capturing the recent  star formation activity that might be related with injection of chemical elements and/or the triggering of outflows, which could affect the chemical abundances profiles on the discs in the recent past.

\begin{figure}
\resizebox{8cm}{!}{\includegraphics{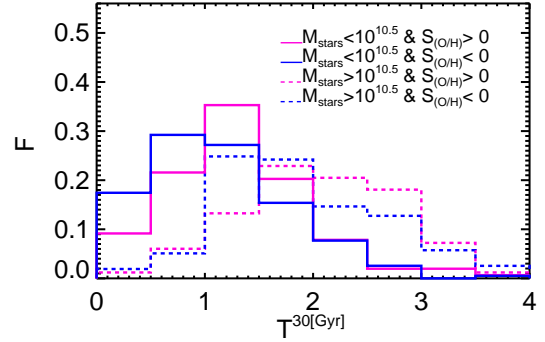}}
\caption{Distributions of lookback times corresponding to the formation of the youngest 30 per cent of stars for high (dashed lines) and low (solid lines) stellar mass galaxies with discs exhibiting positive (magenta lines) and negative (blue lines) gaseous oxygen abundance slopes. Discs with negative  oxygen gradients tend to have more recent star formation  activity. }
\label{HISTage020}
\end{figure}

 Figure \ref{HISTage020} shows the $T^{30}$  distributions for the four analysed subsamples.
As can be seen, high mass galaxies formed their 30 per cent youngest stellar population earlier than  low mass galaxies by   $\sim 1-1.5$ Gyr.
Within each mass subsample, discs with negative metallicity slopes have typically younger stellar populations than those with positive metallicity slopes. The differences are statistically significant according to a KS test, although
they are larger for the low mass galaxies (KS tests: $p=0.09$ and $p=0.19$, with a similar deviation $\approx 0.30$, for the low and high mass subsamples, respectively).
Hence, positive metallicity gradients are associated with discs that have more active star formation  $\approx 1.5 $ Gyr ago, on average, while negative metallicity
slopes tend to be related with  galaxies that have  more recent star formation activity.  

The stellar and AGN feedback play an important role in the regulation of the star formation activity by heating the  dense gas clouds, quenching star formation.
 Additionally, metal mass-loaded outflows can be triggered, removing material from the galaxies that can modify the chemical patterns. Considering that oxygen is released on short time-scales ($\sim 10^7$yr), significant fractions of the $\alpha$-elements could have been transported outwards by galactic winds \citep{oppenheimer2017,machado2018}. If the feedback mechanisms eject material preferentially from the inner regions of galaxies then,  the mean central oxygen abundances should be lower in discs with positive \soh than
those   with negative \sohe. Even more, if the ejection of enriched material in the central regions were the dominant mechanism, we would expect  the outer regions of
the discs to have the same level of enrichment  regardless of \soh (i.e. positive or negative \sohe). Secular evolution will also work in the same direction by driving inwards less-enriched gas  from the outer regions of the discs, diluting the central metallicities \citep[e.g.][]{dimatteo2009, perez2011,sillero2017}.
Conversely, similar central abundances and different levels of enrichment in the outer parts
 would imply different gas accretion histories in the outskirts,  with significant contribution of enriched material in those discs with positive slopes at $z=0$. 

\begin{figure}
\resizebox{8cm}{!}{\includegraphics{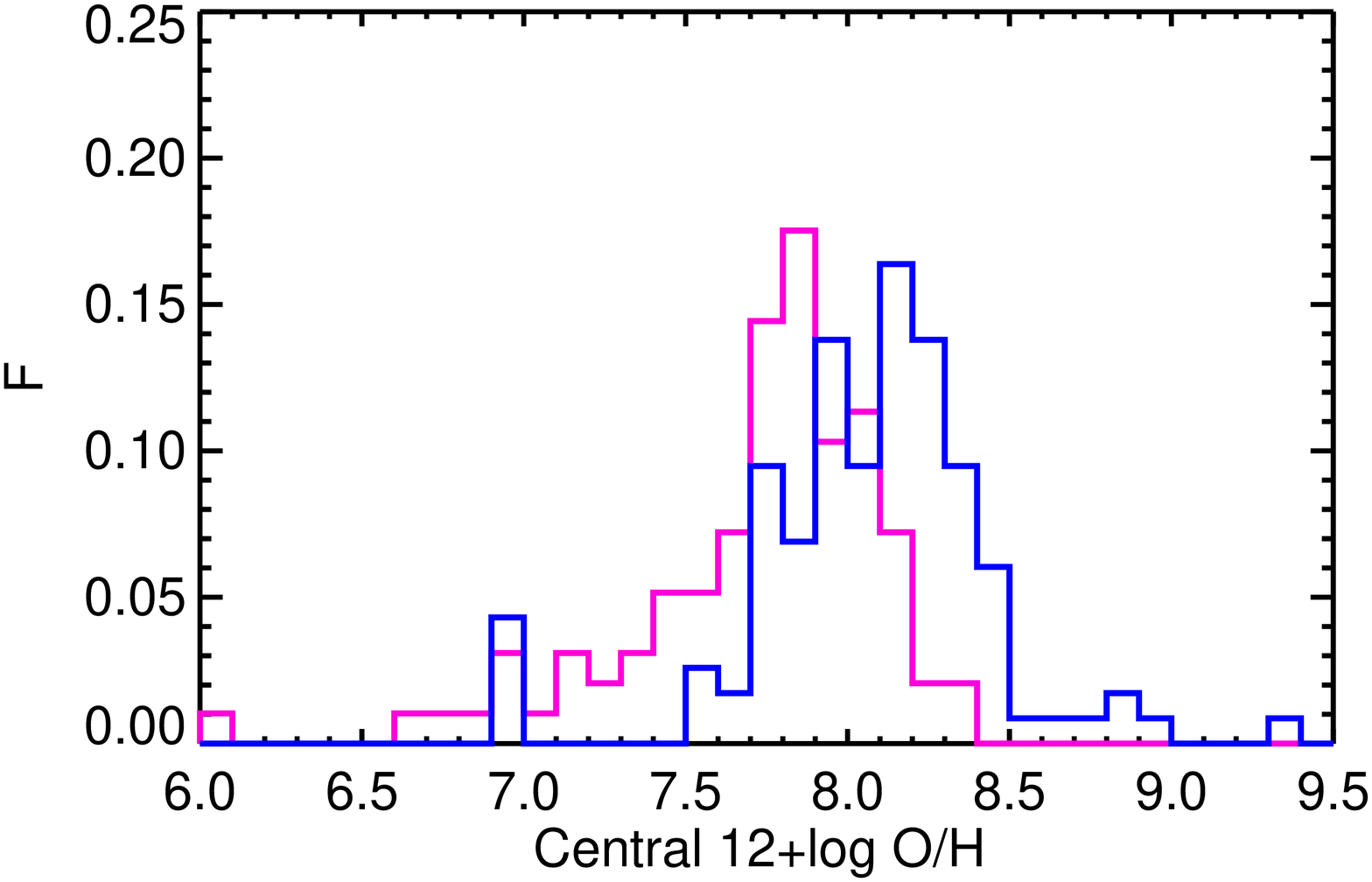}}
\resizebox{8cm}{!}{\includegraphics{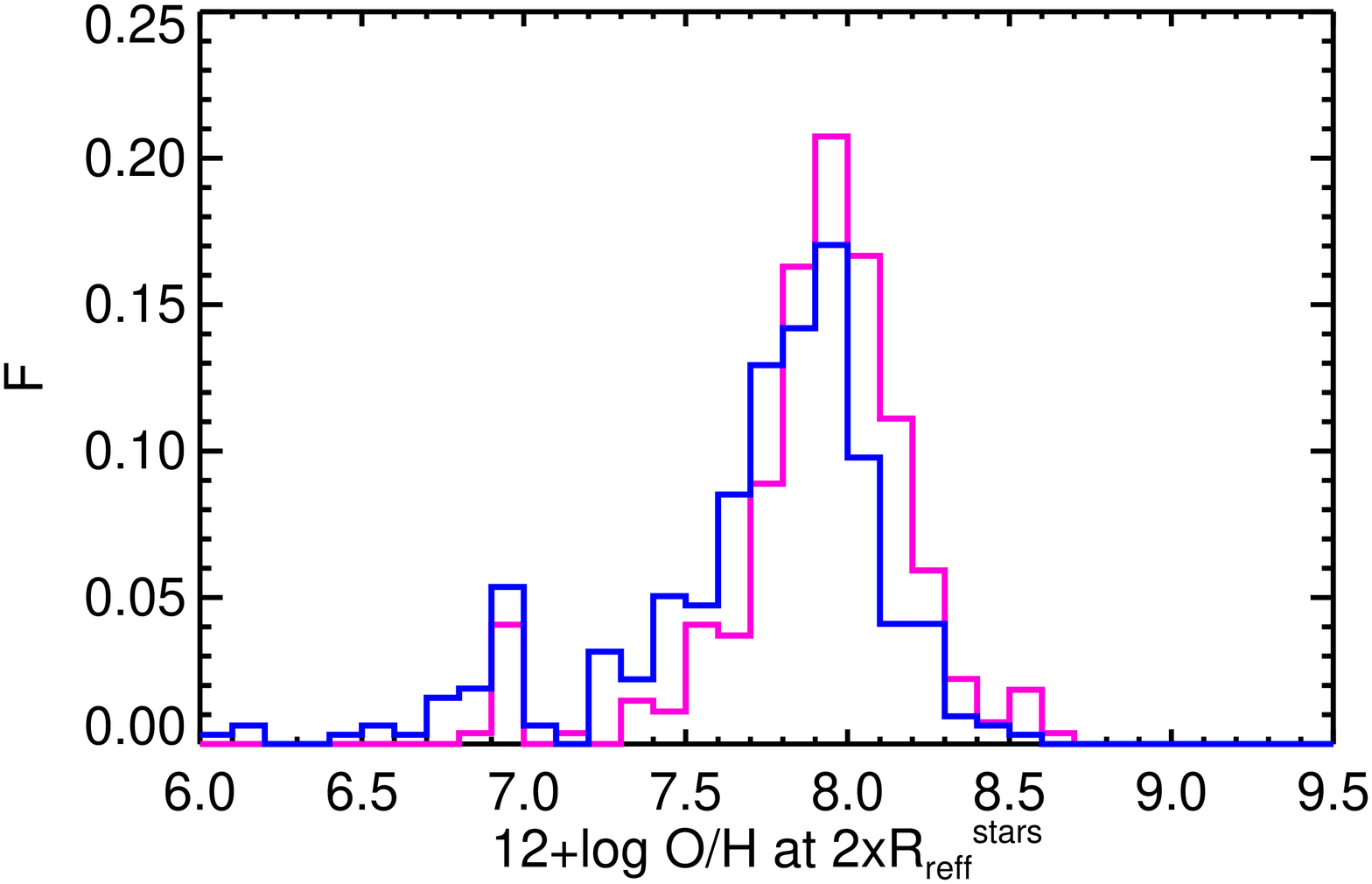}}
\caption{Central gaseous oxygen abundance  for discs with positive (magenta, solid lines) and negative (blue dashed lines) metallicity profiles (upper panel) and the gaseous oxygen abundance at twice the stellar half-mass radius of the discs (lower panel).}
\label{histgrad_innerouter}
\end{figure}

Therefore, to assess if the metallicity gradients are consistent with a decrease of the central metallicity (independently of the physical mechanisms responsible for such  effect), in the upper panel of Fig.~\ref{histgrad_innerouter}, we show the central oxygen abundance 
for discs with positive (magenta, dashed lines) and negative (blue, solid lines) \sohe. The lower panel shows the oxygen abundance at twice the stellar disc effective radius. It is clear that the metallicities of the gas in the disc outskirts are similar for both types of systems. However, discs with positive abundance gradients have systematically lower level of enrichment in the central regions. As mentioned above, the opposite trend would be expected if positive gradients were built mainly as a result of  significant accretion of enriched material in the disc outskirts. This suggests the action of either secular evolution and/or  the stellar and AGN feedback to remove enriched gas from the
central regions. The analysis of the EAGLE set of simulations with the same initial conditions but  with different SN feedback energy yield a trend to have more
positive  slopes for higher SN energy released per event in low mass galaxies (see Appendix A.2).

\begin{figure}
\resizebox{8cm}{!}{\includegraphics{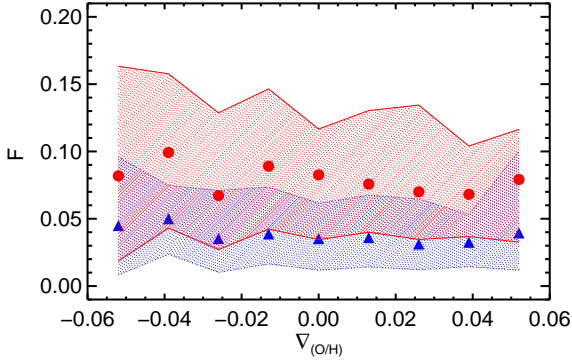}}
\caption{Median fraction of the stellar mass that was formed from  directly heated and enriched gas that, later on, was re-accreted to contribute to the formation of the inner (red, circles) and outer (blue, triangles)
regions of the discs as a function of the oxygen abundance gradients. The shaded areas encompasses the 25 and 75 percentile. }
\label{heatedup}
\end{figure}

For the purpose of estimating the possible contribution of re-accreted material to the discs that was first directly heated by feedback mechanisms,
 we  search for the disc stars formed from gas that was previously heated up by more than
$10^7$K  (see Section 2). In Fig.~\ref{heatedup}, we show the fractions of disc stars formed from heated gas particles, located  in the inner and outer
regions  of the discs as a function of \sohe. The inner and outer regions have been defined by using the $R_{\rm eff}^{\rm star}$ as a threshold. 
As can be seen from this figure, the contribution of re-heated gas is larger in the central regions than in the outskirts of discs. There is
a slight trend to have large fraction in the central regions of discs with negative metallicity gradients. This heated and enriched gas contributes
to reinforce the formation of negative metallicity gradients.
In the outer  regions, the fractions are less than five per cent. Hence, the re-accretion of enriched material ejected by feedback mechanisms is not  contributing significantly in the outer parts of the discs, supporting the conclusions drawn from Fig.~\ref{histgrad_innerouter}.

\begin{figure}
\resizebox{8cm}{!}{\includegraphics{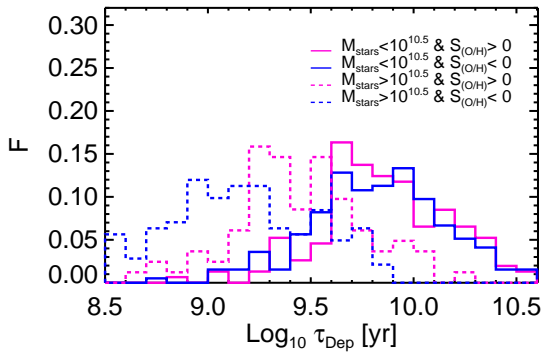}}
\caption{Distributions of the depletion times estimated using the HI and H2 gas mass estimated by  \citet{lagos2015}  for high (dashed lines) and low (solid lines) stellar mass galaxies with discs exhibiting positive (magenta lines) and negative (blue lines) oxygen abundance slopes. }
\label{depletion}
\end{figure}

The gas depletion times ($\tau_{\rm Dep}= {\rm M(HI)+M(H2)/SFR}$) provide an estimate of the importance of the star formation activity in relation to the gas availability of the system.
To estimate them, we use the HI and H2 masses (M(HI)+M(H2)) calculated by \citet{lagos2015} by applying the model of \citet{krumholz2013}\footnote{Similar relative trends
are found if the star-forming gas is used to estimate the depletion times.}. As can be seen 
from Fig.~\ref{depletion}, massive galaxies have shorter neutral gas depletion times than lower stellar mass galaxies as a result of their higher gas surface densities. 
Interestingly, within each mass interval, galaxies with negative \soh tend to have longer depletion times (KS tests: $0.43,~p=0.18$ and $0.9,~p=0.22$ for 
the low and high mass subsamples, respectively). This is consistent  with the fact
that these galaxies have been able to extend their star formation activity to more recent times. We note that the differences are stronger for more massive galaxies.

From the previous analysis, we find that discs with negative \soh tend to have formed stars more actively in the recent past, have slightly higher contribution of
stars formed from heated gas by feedback in the central disc regions and longer depletion times than those with positive \sohe. The comparison of the inner and outer levels of enrichments of 
discs with positive and negative \soh is consistent with the action of mechanisms that diluted the central abundances in discs with positive \soh.

\begin{figure}
\resizebox{8cm}{!}{\includegraphics{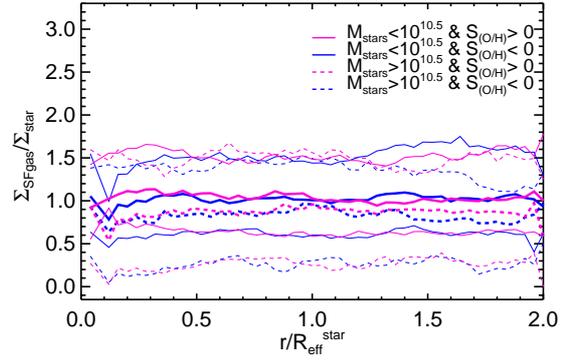}}
\caption{ Median ratio between the gas surface density $\Sigma_{\rm SFgas}$  and the young stellar surface density $\Sigma_{\rm star}$ as a function of $r/R^{\rm star}_{\rm eff}$ for EAGLE discs with  positive (magenta lines) and negative (blue lines) gaseous oxygen abundance slopes. The profiles are shown for  high (dashed lines) and low (solid lines) mass galaxies. The thin lines
denote the first and third quartiles.   }
\label{sigma}
\end{figure}

As clearly shown in Fig.~\ref{gradients}, the EAGLE discs have  median negative gradients that do not  vary significantly  as a function of stellar mass when all galaxies in the analysed sample are considered. However,  they are shallower than observed. To understand the origin of this overall weak median metallicity gradient (see  the lower panel of Fig. 2), in Fig.~\ref{sigma}  we show the ratio of the star-forming gas surface density ($\Sigma_{\rm SFgas}$) to the stellar surface density of stars younger than 2 Gyrs ($\Sigma_{\rm star}$) for discs with high and low stellar mass and within each subsample, for those with negative and positive metallicity gradients, as a function of  $r/R^{\rm star}_{\rm eff}$.
As can be seen the median ratios are close to unity as a function of radius, implying similarly high and similar efficiency to transform the available gas into stars regardless of the location of the star-forming regions in the discs.
This trend suggests that the oxygen content in the ISM is built up quickly, resulting in flat oxygen profiles \citep[e.g.][]{pilkington2012,molla2017}. This efficient
star formation activity provides an interpretation to  the overall weak (negative/positive) metallicity gradients found in the EAGLE discs. 

It is important to note that there is a large spread in the  star forming gas-to-stellar surface density ratios that accounts for
a wide variety of individual behaviours determined by each evolutionary history, as can be seen from   Fig.~\ref{sigma}  (thin lines denote the first and third quartiles). 

\subsection{The role of mergers and the environment}

\begin{figure*}
\resizebox{12cm}{!}{\includegraphics{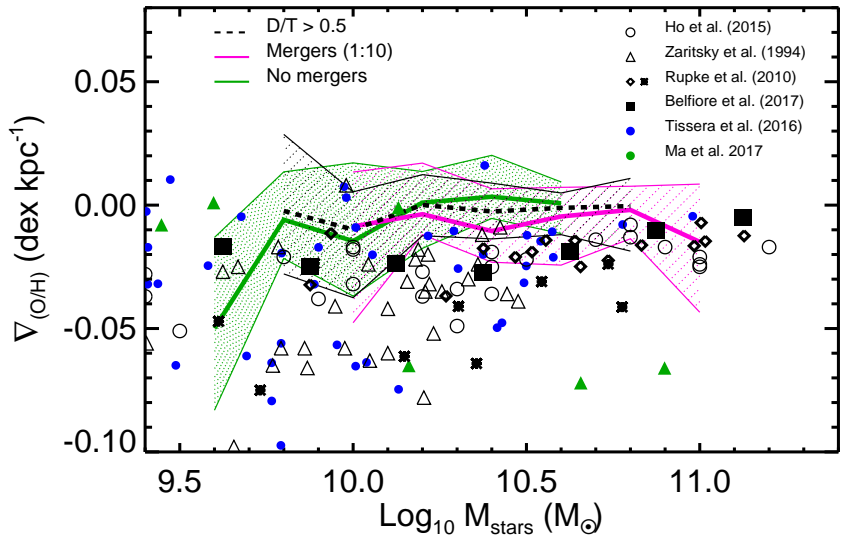}}
\caption{Median abundance gradients estimated for the star-forming disc components as a function of the stellar mass of the EAGLE galaxies. Galaxies which experienced no mergers determine a correlation between the oxygen abundance gradients and the stellar mass (green lines). Those which have had a merger (more massive than 1:10) are 
displayed separately and determine a flat relation (magenta line). The relation determined for galaxies with D/T$ >0.5$ and total gas fraction larger
than 0.15 is also shown for comparison (black line). The shaded areas are defined by the first and third quartiles.
 For comparison observational results from  \citet[][open triangles]{zaritsky1994}, \citet[][black diamonds]{rupke2010}, \citet[][open circles]{ho2015} and
\citet[][fillerd squares]{belfiore2017} are included. We also include the simulated results by \citet[][blue solid circles]{tissera2016}  and \citet[][green filled triangles]{ma2017}  for metallicity gradients of discs simulated with different numerical codes and subgrid physics.}
\label{mergers}
\end{figure*}

For the selected EAGLE galaxies, we  estimate the mass and time of  merger events with relative stellar mass ratios larger than  1:10.
  In Fig.~\ref{mergers},
we show \soh as a function of stellar masses for galaxies that had  merger events and those that did not (hereafter, non-merger galaxies).
Galaxies with quiet assembly histories   show a clearer trend with stellar mass. This trend is robust against numerical resolution as can be seen from Appendix A.1, where 
we show results for the high-resolution \RECAL~simulation.

In  Fig. ~\ref{mergers},  we also display the relation for those galaxies with D/T$ >0.5$ and  gas fraction larger than 15 per cent (dashed line). These galaxies follow the relation for non-merger galaxies and are more comparable to spiral galaxies for which observational results are commonly reported  \citep[][and references therein]{ho2015}. However, we note that the stellar mass range
analysed  in observations extends to lower stellar masses which are not well-resolved in the \REF~simulation.  To extend the analysis to lower masses, a similar
study is performed on the \RECAL, finding that the simulated trend  continues down to $\sim 10^{9.2}{\rm M_{\sun}}$ (see Appendix A.1). 
If a  stellar mass threshold of $10^{10}{\rm M_{\sun}}$ is assumed, we estimate  median \soh of $-0.010 \pm 0.002$ and $-0.015 \pm 0.005$ for the high and low stellar-mass subsamples, respectively. The change of the slopes is consistent with observational results in the sense that smaller galaxies tend to have more negative gradients, on average, although the  difference in simulations is marginal.

For comparison with other simulations, in Fig. ~\ref{mergers}, we have also included the \soh reported by \citet{tissera2016} and \citet{ma2017}. Both works find steeper negative metallicity gradients. The main differences between these works originate in the different subgrid physics that  affects the regulation of the star formation activity and the mixing of chemical elements (see Introduction).

\begin{figure}
\resizebox{8.5cm}{!}{\includegraphics{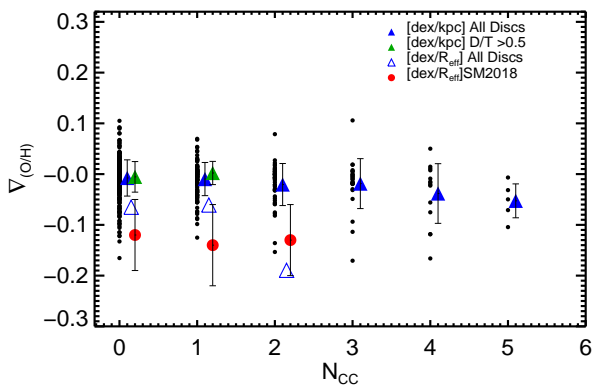}}
\caption{Median oxygen abundance gradients (dex~kpc$^{-1}$) as a function of $N_{\rm CC}$ for all analysed discs (blue triangles), for discs with D/T$>0.5$ (green triangles). Median normalised gradients 
(dex$/R^{\rm star}_{\rm eff}$) are also included (open blue triangles). Individual gradients for the simulated discs are shown (small black circles). For comparison, the mean normalised metallicity gradients reported by \citet{sanchezMen2018} are also included (red filled circles; MS2018).}
\label{environ}
\end{figure}

In Fig.~\ref{environ} we show \soh as a function of the environment. The value N$_{\rm CC} =0$ corresponds to isolated galaxies
within the 500 kpc cube volume, according to the adopted criteria discussed in  Section 2. As can be seen from this figure, there are no clear variations with  local environment.
The small decrease towards more negative metallicity gradients for larger groups is produced by a larger contribution of more massive dispersion-dominated galaxies located in  denser environments.
These galaxies  have small-size disc components (Fig.4). The variations between the median gradients as a function of environment are within
the standard deviations.  

The median values for galaxies with $D/T >0.5 $ are  included in Fig.~\ref{environ} (green triangles). They also show no trend with N$_{\rm CC}$.
To better compare with the recent observational findings by \citet{sanchezMen2018}, the normalised metallicity gradients are additionally shown (open blue squares).
Our results agree with observations that reports no clear effects associated to variation in the global  environment (red solid circles).
 However, as mentioned before, a larger galaxy sample covering the same mass range in different environments is required to improve this analysis.

\subsection{Normalised oxygen  gradients}

Observational results suggest the existence of a characteristic metallicity gradient when the abundance profiles are normalised
by a scale-length related to the size of the galaxies \citep[e.g.][]{sanchez2013Califa}. Because the size of galaxies correlates with stellar mass, the
renormalisation of the metallicity profiles by the effective radius (dex r$_{\rm eff}^{-1}$) removes such a trend.

  As discussed  above, on the one hand, 
 \soh  shows no clear trend with stellar mass, except for non-merger galaxies that have weaker \soh for more massive galaxies. On the other hand, the relation between galaxy size and stellar mass is well-reproduced by the EAGLE galaxies \citep{schaye2015}.
The stellar disc scale-lengths  R$_{\rm eff}^{\rm star}$ show a correlation with stellar mass albeit slightly weaker (Tissera et al. 2018, in preparation) than that reported for CALIFA disc galaxies by \citet{sanchezb2014}. 

\begin{figure}
\resizebox{8cm}{!}{\includegraphics{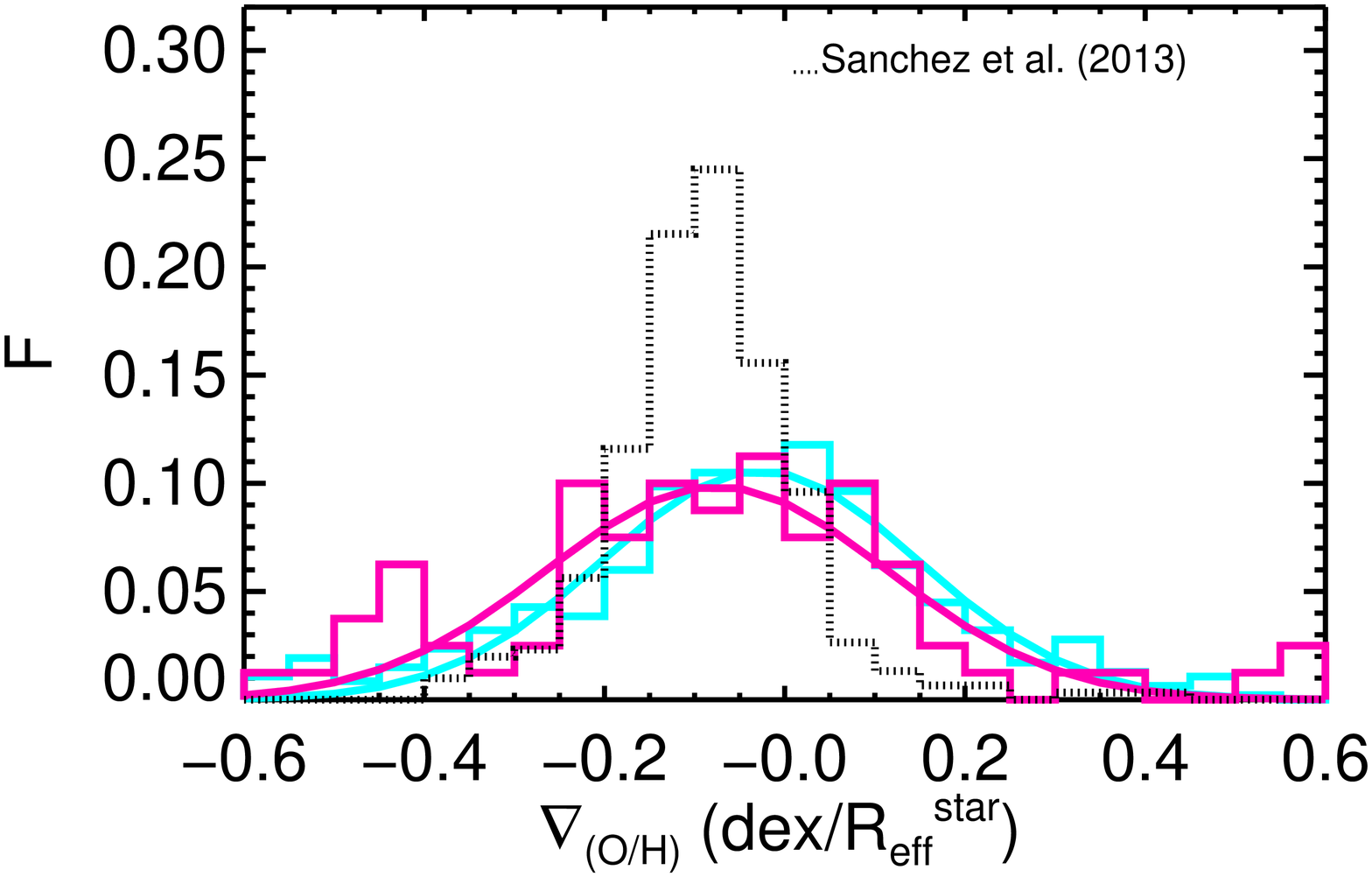}}
\resizebox{8cm}{!}{\includegraphics{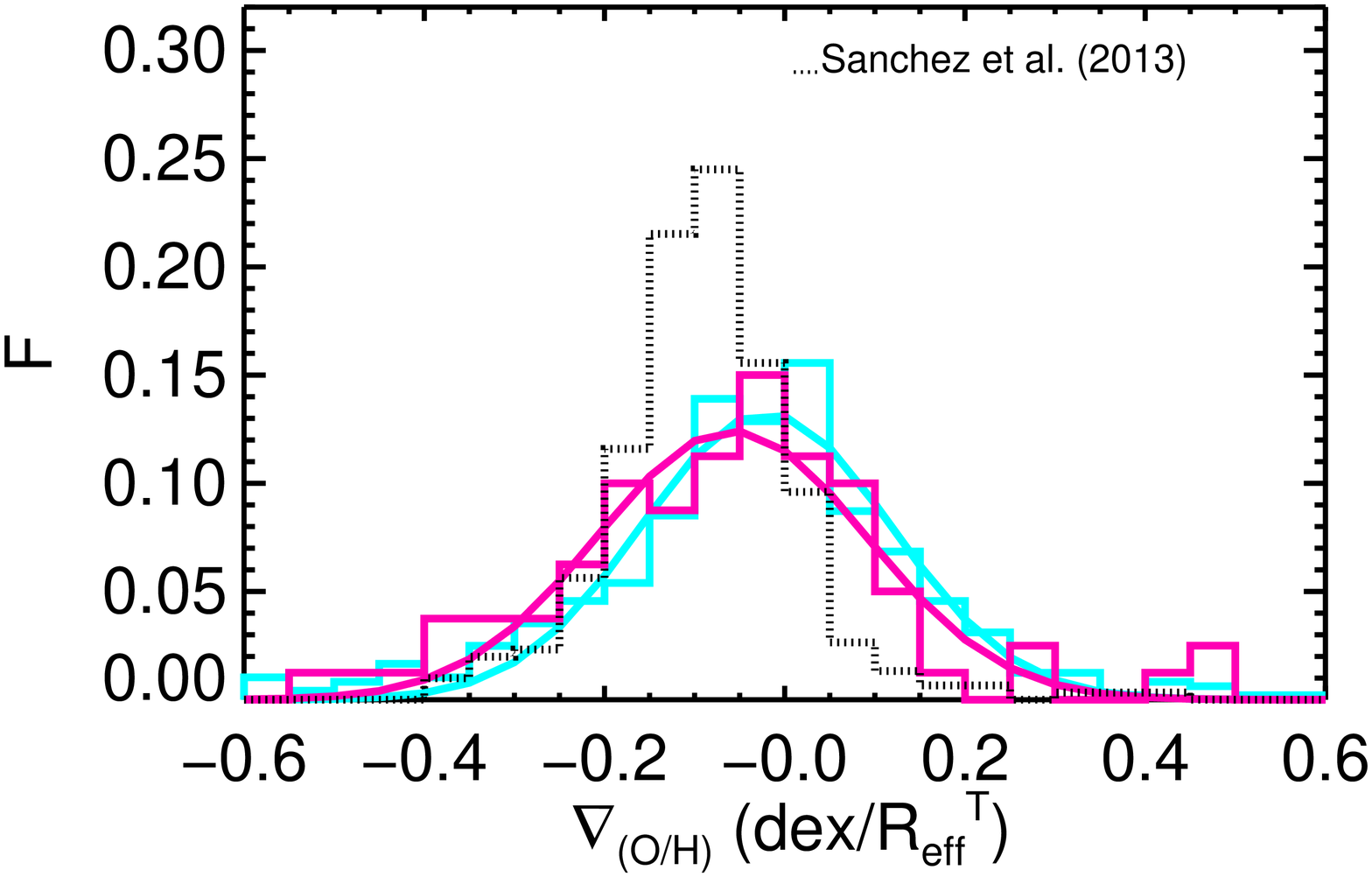}}
\caption{Distribution of the normalised metallicity gradients of the star-forming gas in  the analysed EAGLE discs  by using the R$_{\rm eff}^{\rm star}$ (upper panel) and R$_{\rm eff}^{\rm T}$ (lower panel),   in the two defined environments (low density ($N_{\rm CC} \leq 2$): cyan lines and high density  ($N_{\rm CC} > 2$: magenta lines)
as given in Fig.2. The two solid lines represent the best gaussian fits to each distribution. For comparison, the observational distribution of normalised slopes reported by \citet{sanchez2013Califa} is included (black, dashed line).
}
\label{histonorm}
\end{figure}

We estimate the normalised \soh
within the same radial range $[0.5,2] R_{\rm  eff}^{\rm star}$ used in the previous section. The normalisation has been also done by using  $R_{\rm  eff}^{\rm T}$.   Fig.~\ref{histonorm}  shows the distributions of the normalised  \soh for the two
defined environments (we applied the same criteria explained in Section 2). As can be seen from the figure,  the normalised oxygen gradients are in general shallower than those reported for the CALIFA galaxies
 by \citet{sanchez2013Califa}.  
In both panels, we include the best Gaussian fits to the \soh distributions in low-and-high density regions. As can be seen,  distributions  are slightly displaced towards more positive values for discs in low-density regions (see Table~\ref{table2} for a summary of the fitting parameters). 
The gaussian fits to the total
distributions yield centred values at  $-0.03$  dexR$_{\rm  eff}^{-1}$ for both of them. However the  standard deviations are narrower when the abundance profiles are normalised by R$_{\rm eff}^{\rm T}$.

\begin{figure}
\resizebox{8cm}{!}{\includegraphics{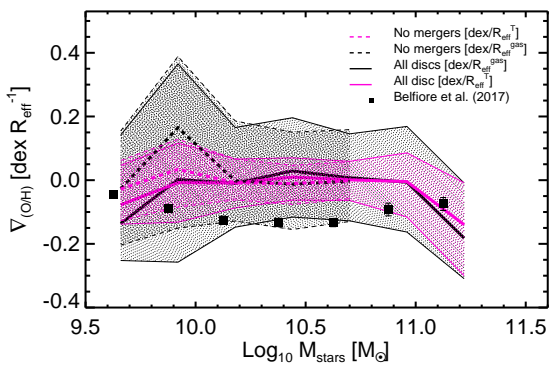}}
\caption{Median oxygen gradients of the star-forming gas in discs  as a function of stellar mass normalised by R$_{\rm eff}^{\rm star}$ (black lines) and R$_{\rm eff}^{\rm T}$ (magenta lines). Similar relations for galaxies with no mergers are also included (dashed lines). For comparison, the observational results from \citet{belfiore2017} are included (black squares).
}
\label{scatter_recal}
\end{figure}

In  Fig. ~\ref{scatter_recal}, the normalised gradients as a function of stellar mass are displayed. As can be seen from this figure there are no clear trend with stellar mass in agreement with observations. The scatter of \soh normalised by  R$_{\rm eff}^{\rm T}$ is significant smaller as mentioned before.
This is expected considering that there is a clearer correlation between this scale-length and galaxy stellar mass.
Similar behaviours are found for   non-merger galaxies. In agreement with observational results, the relation between \soh and stellar mass 
is no longer present in the normalised gradients.

\begin{table}
\caption{Parameters of the Gaussian fits to the normalised \soh distributions of galaxy discs in low ($N_{\rm CC}\leq 2$) and high $N_{\rm CC } > 2$ density regions.  }
\begin{tabular}{|l|c|c|c|c|c|}
\hline
\soh & $N_{\rm CC}$               &  Height           &    \soh$^{\rm central}$         & $\sigma$ &  $\chi^2$  \\
\hline
\multirow{3}{*}{ dex/R$_{\rm  eff}^{\rm star}$}&$N_{\rm CC} >2$ & 0.10  &  $-0.07$& 0.19 &0.0004 \\
& $N_{\rm CC}\leq 2$ & 0.11  &  $-0.03$& 0.17&0.0004\\
& All  & 0.11  &  $-0.03$& 0.18 &0.0001\\
\multirow{3}{*}{ dex/R$_{\rm  eff}^{\rm T}$}&$N_{\rm CC} > 2$ & 0.12  &  $-0.06$& 0.14 &0.0003 \\
& $N_{\rm CC} \leq 2$  & 0.13  &  $-0.02$& 0.14&0.0001\\
& All  & 0.13  &  $-0.03$& 0.14 &0.0001\\
\hline
\end{tabular}
\label{table2}
\end{table}

\section{Conclusions}

We analysed the oxygen abundance gradients of the star-forming gas in  discs of 592 central galaxies, identified in the \REF~ simulation of the EAGLE project.
We studied them as a function of stellar masses, recent
merger history, gas richness and global environment. We also analysed the high resolution \RECAL~ and a suite of smaller volume runs with different subgrid physics (see Appendix).

Our main results can be summarised as follows:
\begin{itemize}
\item The median oxygen gradients of the EAGLE discs are negative,$-0.011 \pm 0.002$ dex kpc$^{-1}$, albeit shallower than observed.
The negative slopes are a consequence of the global inside-out disc assembly. 
We find no clear trend of \soh  with stellar mass (Fig.~\ref{gradients}).

\item The  \soh  is found to correlate with the scale-length of the gas discs so that  shallower slopes are measured in larger discs (Fig.~\ref{gradients_reff}).  Small 
gas discs show a larger variety of \soh with most disc negative gradients in dispersion-dominated galaxies (Rosito et al., in preparation). The correlation is found to be mainly
determined  by discs with negative \soh while  discs with positive \soh show no trend with  gas scale-length.

\item There is a significant fraction of galaxies  with  positive disc \soh  ($\sim 40$ per cent).  These galaxies show  important star formation activity $\approx~1.5 $ Gyr before those with negative gradients at
a similar stellar mass (Fig.~\ref{HISTage020}). 
Analysing the gas abundances of the inner and outer parts of discs, we find evidence suggesting the action of the stellar/AGN feedbacks that
preferentially removes metal-rich gas from the central regions (Fig.~\ref{histgrad_innerouter}). Secular evolution may  also contribute by diluting the central abundances.
There is a weak trend for galaxies with positive \soh to prefer low-density regions (Fig.~\ref{dt_environ} and Fig.~\ref{histonorm}). 

\item The contribution to the formation of the discs and the abundance gradients of  material previously heated up by stellar/AGN feedback, and later on re-accreted, is  small in the analysed discs (below 10 per cent, on average). However,  we detect slightly larger contributions of re-accreted material
in the central regions of discs with negative \soh  than in those with positive ones. In the outer regions of discs, this contribution is  smaller (about 5 per cent on average)  and similar for all 
discs, regardless of their oxygen slopes (Fig.~\ref{heatedup}).

\item The   weak metallicity  gradients measured in the EAGLE discs can be ascribed to the overall high star formation efficiency detected in the discs (regardles of radius (Fig.~\ref{sigma}).  
The gas with physical conditions suitable for forming stars is transformed efficiently
along the discs. Therefore, oxygen profiles converge quickly to a weak metallicity gradient, on average. However, there is significant dispersion that reflects the large variation in the star formation histories of the discs.

\item Galaxies with  quiet merger histories 
show  a positive trend between \soh and   stellar mass so that lower mass galaxies tend to have steeper metallicity gradients, on average (Fig.~\ref{mergers}). Non-merger galaxies  tend to be rotational-dominated (D/T$>0.5)$. However, at low stellar masses, there is a
significant contribution of intermediate morphologies ($0.3< {\rm D/T} <0.5)$. These galaxies tend to be  located in low-density environments.
The normalisation of the oxygen profiles by the disc scale-length tends to remove the trend, in agreement with observations (Fig.~\ref{scatter_recal}).
These findings suggest that the large variety of gas-phase oxygen gradients at a given stellar mass reported by observations might originate in the fact that galaxies with different assembly histories  are considered together.


\item No clear  differences in the  gas disc \soh are found between  galaxies located in groups and in isolation, in agreement with recent observational results (Fig.~\ref{environ}).
However, we note that the analysed disc sample does not cover the same mass range in all environments. Hence, a larger and higher resolution galaxy sample is needed to analyse this effect in more detail.
 

\end{itemize}

\section*{Acknowledgments}
{RAC is a Royal Society University Research Fellow.
CL is funded by a Discovery Early Career Researcher Award
(DE150100618) and by the Australian Research Council Centre of
Excellence for All Sky Astrophysics in 3 Dimensions (ASTRO 3D),
through project number CE170100013. MS acknowledges the VENI grant 639.041.74.
PBT acknowledges partial funding by Fondecyt Regular 2015 - 1150334.
This project has received funding from the European Union Horizon 2020
Research and Innovation Programme under the Marie Sklodowska-Curie
grant agreement No 734374.

This work used the DiRAC Data Centric system at Durham University, operated by the Institute for Computational Cosmology on behalf of the STFC DiRAC HPC Facility (www.dirac.ac.uk). This equipment was funded by BIS National E-infrastructure capital grant ST/K00042X/1, STFC capital grants ST/H008519/1 and ST/K00087X/1, STFC DiRAC Operations grant ST/K003267/1 and Durham University. DiRAC is part of the National E-Infrastructure. We acknowledge PRACE for awarding us access to the Curie machine based in France at TGCC, CEA, Bruyeres-le-Chatel.
}
\bibliographystyle{mnras}
\bibliography{bibliography}

\appendix
\section{Assessment of numerical resolution effects}
{
The analysis presented in this work is based on the galaxies identified in the 100 Mpc cubic volume (\REF). The selected sample comprises gas discs with  a large variety of \sohe. 
As discussed in Section 2, a minimum number of particles has been adopted to diminish resolution problems in the estimation of the metallicity profiles.  In order to further check the effects of resolution, we also estimate the \soh
in the high resolution run of 25 Mpc box size,  \RECAL~ (see Table 1). This simulation has approximately an order of magnitude higher numerical resolution than \REF, allowing us to
both check the robustness of the results against resolution and explore if smaller galaxies  (M$< 10^{9.5}{\rm M_{\odot}}$) follow the same trends.

We identify and measure the oxygen profiles of the disc components in \RECAL, applying the same criteria used for  \REF, therefore extending the mass range down to $\sim10^{9}$ M$\sun$. In Fig.~\ref{mergerscalibrated}, we show \soh 
 as a function of stellar mass (black line). The gradients have been rescaled by -0.03 dex~kpc$^{-1}$. This is the difference estimated between the median simulated \soh
and the observed one by \citet{belfiore2017}  at $\approx 10^{10.5}{\rm M_{\sun}}$. The simulated relation shows a trend consistent with observations  in the
sense that high mass galaxies tend to have gas discs with shallower \soh.

To compare  results from \RECAL~with those obtained for \REF, in Fig.~\ref{mergerscalibrated} we  also show the median relation  for galaxies in  \REF~ in   low density  environment (cyan line; the relation has also been rescaled by -0.03 dex~kpc$^{-1}$). To improve the comparison, we estimate the local environment of galaxies in \RECAL, adopting the same stellar mass threshold  used for \REF. We find that all selected  galaxies in \RECAL~ belong to low density regions,  $N_{CC} \leq 2$. Hence, the relations are consistent so that smaller galaxies show more negative metallicity gradients,  although discs in \RECAL~tend to have slightly flatter gradients.

\begin{figure*}
\resizebox{10cm}{!}{\includegraphics{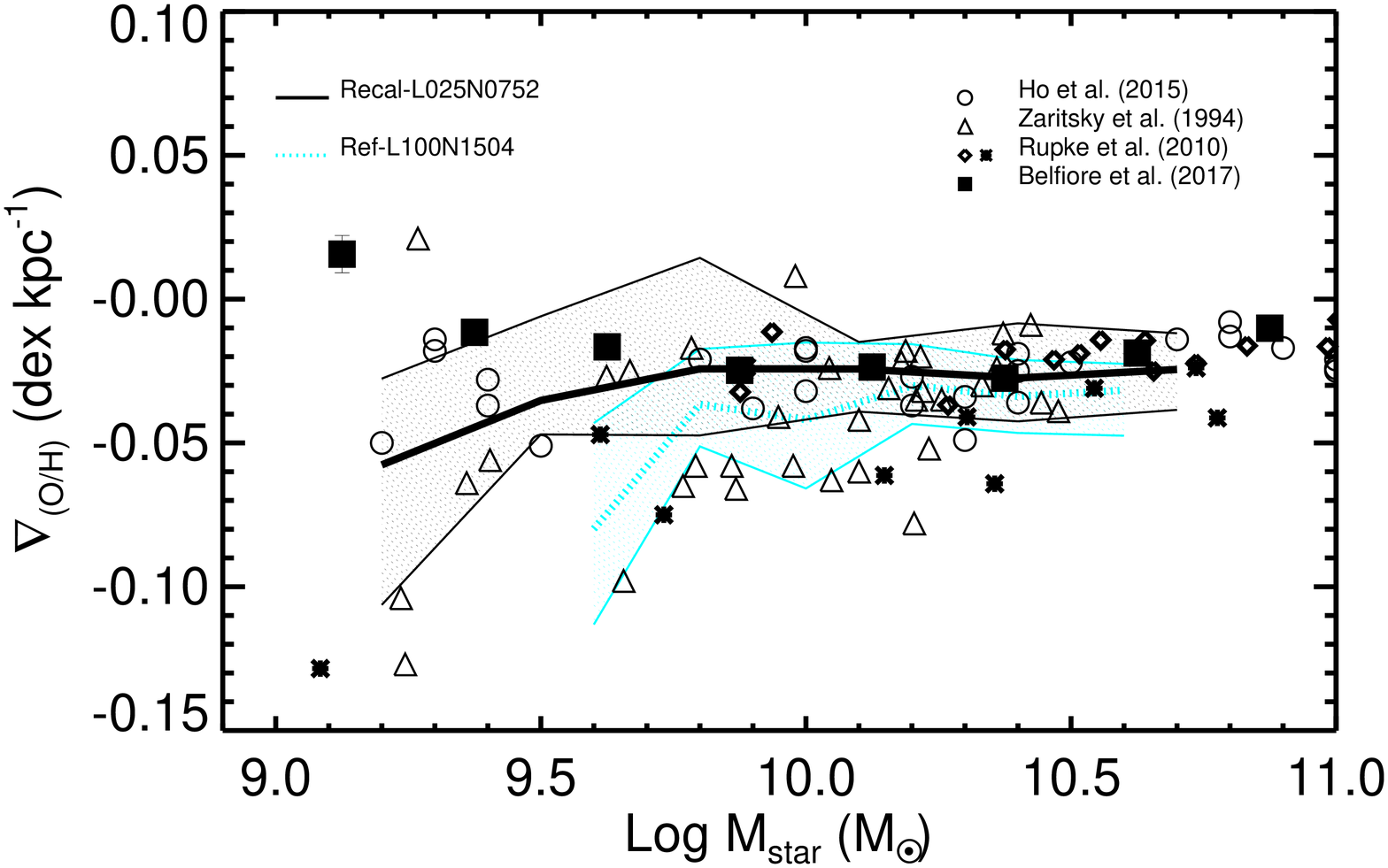}}
\caption{Median \soh  estimated for the star-forming disc components as a function of the stellar mass of the EAGLE galaxies in
for the \RECAL~run (black solid lines). For comparison the simulated relation for the discs in low-density environments of the \REF~ are also shown (cyan lines). The simulated relations have been  rescaled by -0.03 dex~kpc$^{-1}$ to match the mean observed values reported by \citet{belfiore2017} at $\approx 10^{10.5}{\rm M_{\sun}}$. The shaded area are defined by the first and the third quartiles.
}
\label{mergerscalibrated}
\end{figure*} 
}



\section{The effects of SN feedback}

The EAGLE project includes a set of simulations run with the same initial conditions in a 25 Mpc box side volume, adopting different SN feedback energies: Ref-L025N0376, StrongFB-L025N0376 and 
WeakFB-L025N0376 
(see Table 1). This set allows the assessment of  the effects of varying the injected SN energy per event on the galaxy properties and, particularly, on the  oxygen profiles.
The variation of the energy injected by SNe has an impact on the gas fraction as well as on the stellar mass of galaxies as the star formation activity can be  delayed to later times. Additionally, higher SN energy  will also
increase the impact of outflows triggered by starbursts. Both the regulation of the star formation activity and the transport of material out of the galaxy affect the 
chemical abundance patterns of the discs \citep{gibson2013}.

\begin{figure}
 \resizebox{8cm}{!}{\includegraphics{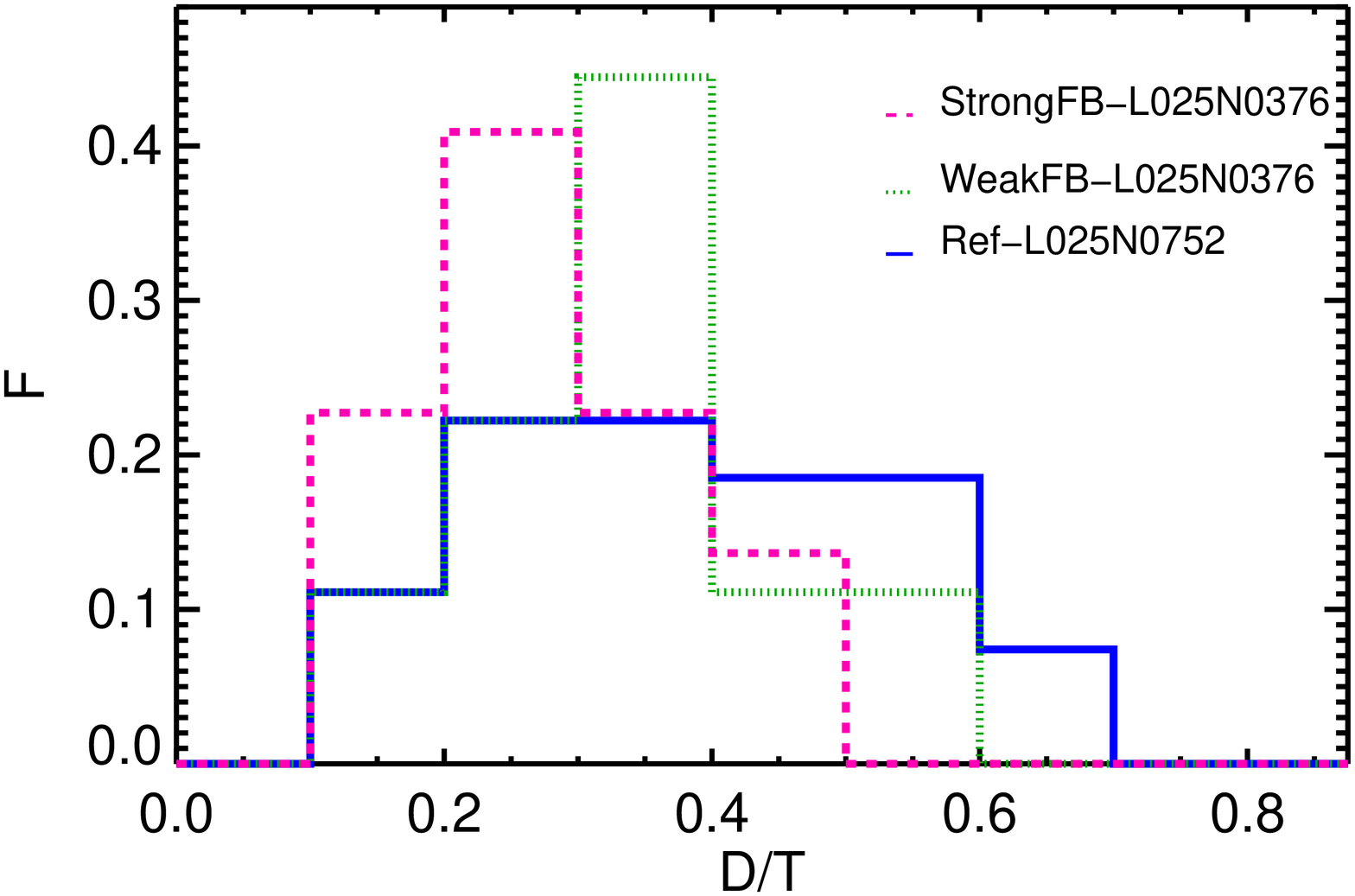}}
 \resizebox{8cm}{!}{\includegraphics{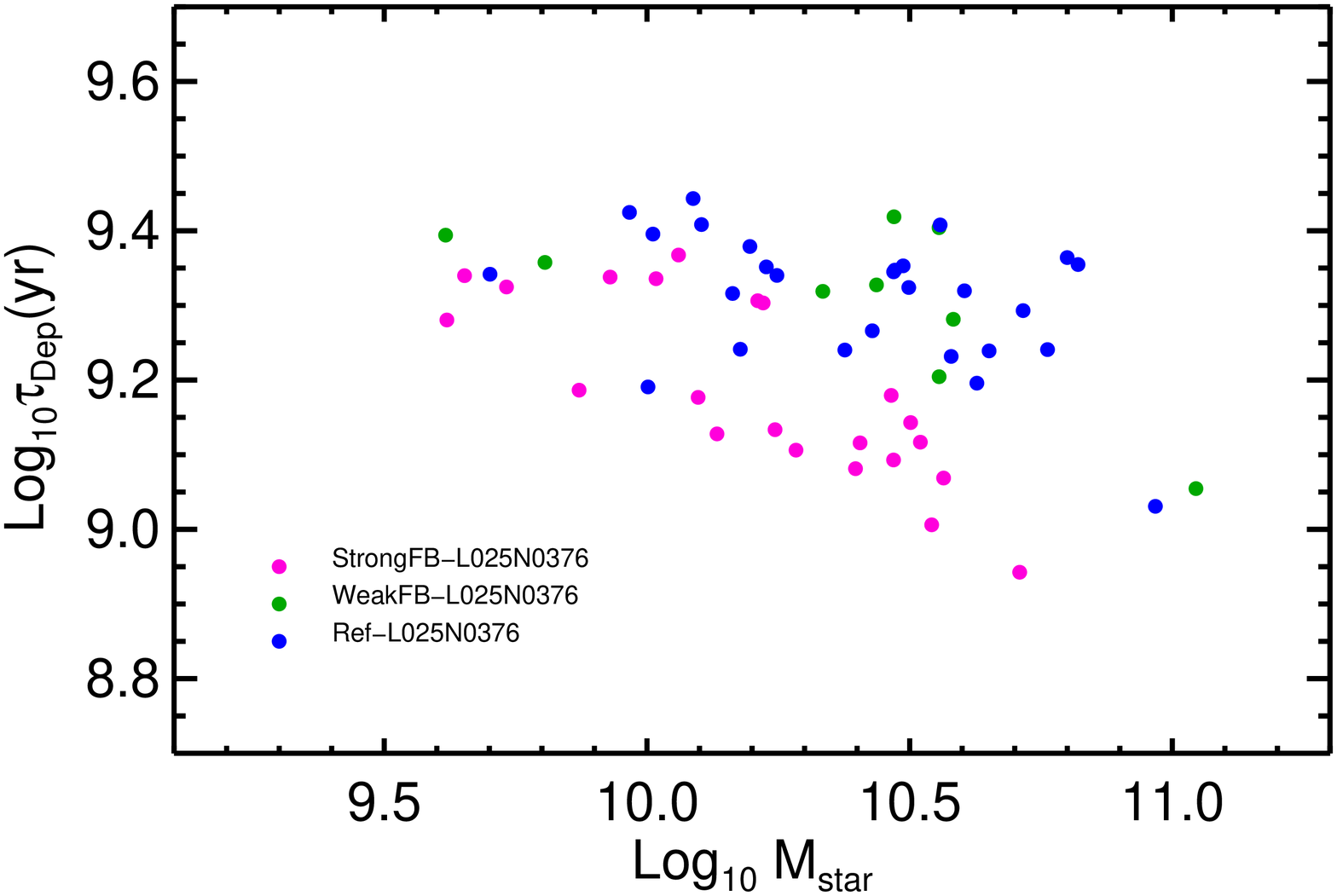}}
 \caption{ Upper panel: Mass-weighted distribution function of the D/T ratios for the analysed galaxies in  Ref-L025N0376 (blue solid line), WeakFB-L025N0376 (green dotted line) and StrongFB-L025N0376 (magenta dashed line). Lower panel: Depletion times as a function of stellar mass for the same set of galaxies.
 }
 \label{morpho25}
 \end{figure}

The D/T ratio can be also affected by variations in the SN feedback. This is relevant for our analysis since we are selecting discs with a minimum number of 1000 baryonic particles and 100 star-forming gas regions  (Section 2). When these conditions are applied,  only 9 galaxies are selected in WeakFB-L025N0376. This is probably the result of an  earlier efficient transformation of gas into stars that did not leave enough material in the discs. In the case of  StrongFB-L025N037,  as reported by \citet{crain2017}, galaxies  have slightly larger HI gas mass at a given stellar mass.
 In Fig. ~\ref{morpho25}, we show the D/T distributions for the three runs. As can be seen, there are
more galaxies dominated by dispersion in the WeakFB-L025N0376 and  StrongFB-L025N0376  than in Ref-L025N0376.

It is also important to note that the galaxies in the  Ref-L025N0376 have total and stellar disc sizes that correlate with  stellar mass in agreement with observations \citep[see also ][]{furlong2015}. However,
the stellar discs in StrongFB-L025N0376 have  sizes with the opposite trends: high mass galaxies tend to have smaller stellar discs by a factor of approximately three compared to those in Ref-L025N0376 and WeakFB-L025N0376.
 
We  estimate the depletion time (i.e. $\tau_{\rm Dep} = {\rm M_{\rm gas}/SFR}$) as a function of stellar mass for the galaxies in the three runs. As can be seen 
from  Fig.~\ref{morpho25},  there is trend for smaller galaxies to have longer depletion times regardless of SN energy adopted for feedback. However,
at a given stellar mass,  galaxies in StrongFB-L025N0376  are currently more efficiently forming stars. This is due to the high star-forming gas density (i.e. the combination of higher gas masses and small disc sizes).

\begin{figure}
 \resizebox{8cm}{!}{\includegraphics{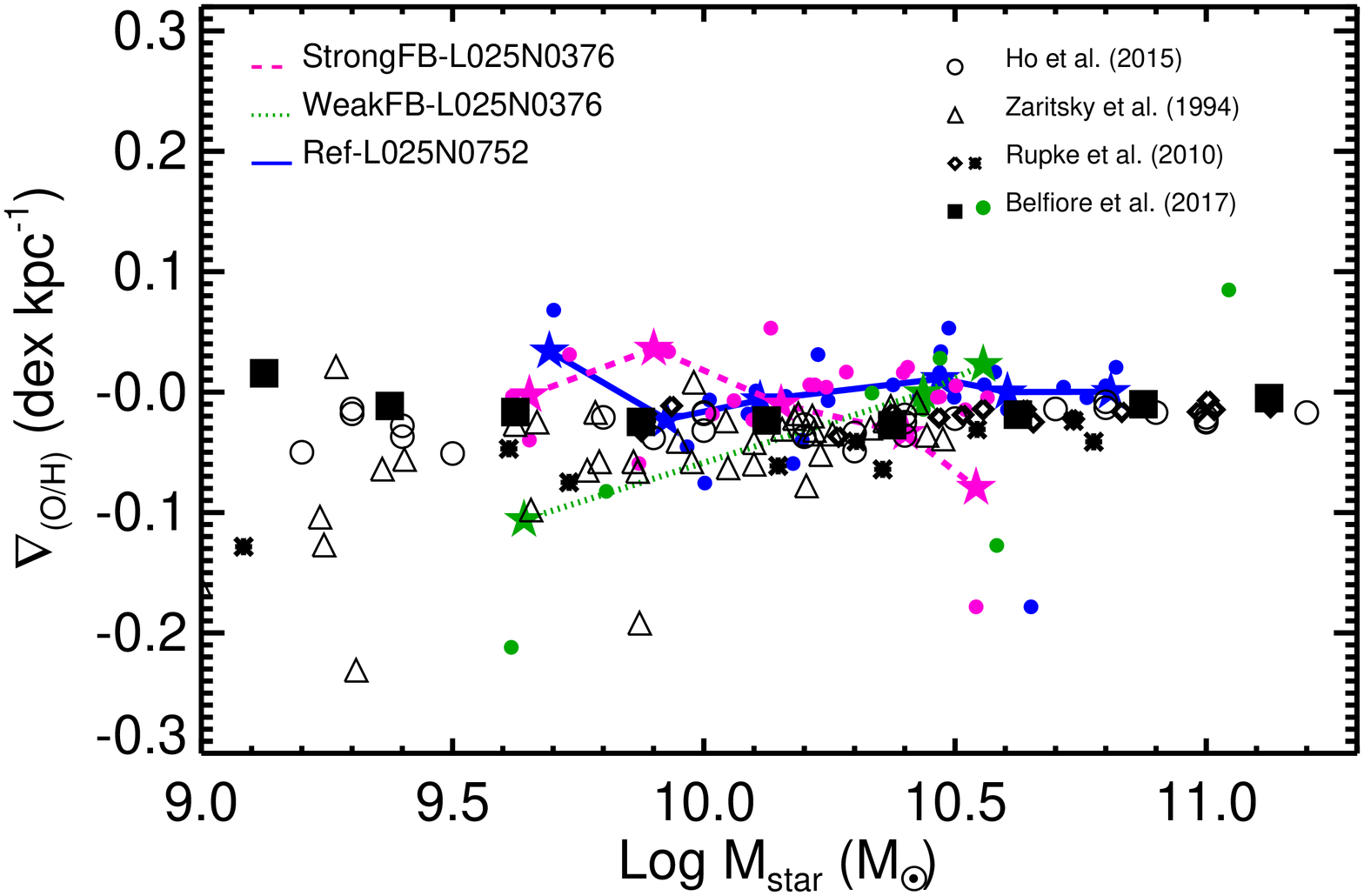}}
 \caption{Median \soh  estimated for the star-forming disc components and the corresponding median values as a function of the stellar mass of the EAGLE galaxies in Ref-L025N0376 (blue solid line and circles), WeakFB-L025N0376 (green line and circles) and StrongFB-L025N0376 (magenta line and circles). For comparison observations have been included as in Fig. 3.
 }
 \label{feedbacks}
 \end{figure}

For the purpose of our work, we are interested in assessing how the \soh change with the SN feedback.  In  Fig.~\ref{feedbacks}, we show the \soh as a function of stellar mass for the three runs.  WeakFB-L025N0376 have 
discs with \soh that  agree well with the observations as in the case of those in Ref-L025N0376. In the case of the StrongFB-L025N0376, there is a change in the relation: more positive \soh  are found in low mass  galaxies (Fig.~\ref{feedbacks}).
The opposite trend can be seen in  WeakFB-L025N0376. However, we acknowledge the fact that the number of galaxies in these samples are small. 

The trends found in this analysis suggest that for enhanced SN feedback, there are more positive \soh in  smaller galaxies, in agreement with previous works \citep{gibson2013}. This indicates that
 SN feecback is playing a role in shaping the relation between \soh and  stellar mass. Stronger feedback can delay the star formation to lower redshift \citep{crain2017}, producing systems with more gas available
to form stars at later times. On the other hand,  smaller galaxies will experience stronger SN driven outflows because of the lower potential well, increasing the possibility to have positive gradients.

\end{document}